\documentclass[pra,aps,amsmath,twocolumn,floatfix,balancelastpage,superscriptaddress]{revtex4-1}

\usepackage{bm}
\usepackage{sidecap} \usepackage[pdftex]{hyperref,graphicx}
\usepackage[usenames, svgnames]{xcolor}
\hypersetup{
 pdftitle = {Revealing quantum statistics with a pair of distant atoms},
 pdfauthor = {C. F. Roos, A. Alberti, D. Meschede, P. Hauke, and H. H\"affner},
	colorlinks=true,linkcolor=blue,citecolor=blue,filecolor=blue,urlcolor=blue,pdfpagemode=UseNone,pdfstartview={XYZ null null 1.00}
}

\usepackage{siunitx}
\usepackage{tocvsec2}

\DeclareMathOperator{\tr}{tr}
\newcommand{\encapsulateMath}[1]{\raisebox{0pt}[0pt][0pt]{#1}}

\newcommand{\figref}[2]{\hyperref[#1]{\getrefnumber{#1}(#2)}}

\usepackage{multirow}
\usepackage{natbib}
\usepackage[active]{srcltx}

\let\origsection\section
\renewcommand{\section}[1]{\textit{#1}.\textemdash}

\renewcommand\textemdash{\leavevmode\unskip\kern0.8pt\rule[0.19\baselineskip]{8pt}{0.4pt}\kern1pt\ignorespaces}

\newcommand{\ket}[1]{\ensuremath{|#1\rangle}}
\newcommand{\bra}[1]{\ensuremath{\langle#1|}}

\newcommand{\spindown}[0]{{\ensuremath{\hspace*{-0.1mm}\downarrow}}}
\newcommand{\spinup}[0]{{\ensuremath{\hspace*{-0.1mm}\uparrow}}}

\begin{document}

\title{Revealing quantum statistics with a pair of distant atoms}

\author{C.~F.~Roos}
\email{christian.roos@uibk.ac.at}
\affiliation{Institut f\"ur Quantenoptik und Quanteninformation
der \"Osterreichischen Akademie der Wissenschaften,
Otto-Hittmair-Platz 1, A-6020 Innsbruck, Austria}
\author{A.~Alberti}
\email{alberti@iap.uni-bonn.de}
\affiliation{Institut f\"ur Angewandte Physik der Universit\"at Bonn, Wegelerstr. 8, 53115 Bonn, Germany}
\author{D.~Meschede}
\affiliation{Institut f\"ur Angewandte Physik der Universit\"at Bonn, Wegelerstr. 8, 53115 Bonn, Germany}
\author{P.~Hauke}
\affiliation{Institut f\"ur Quantenoptik und Quanteninformation
der \"Osterreichischen Akademie der Wissenschaften,
Otto-Hittmair-Platz 1, A-6020 Innsbruck, Austria}
\affiliation{Institut f\"ur Theoretische Physik, Universit\"at Innsbruck, Technikerstra{\ss}e 21a, 6020 Innsbruck, Austria}
\affiliation{Kirchhoff-Institut f\"ur Physik, Ruprecht-Karls-Universit\"at Heidelberg, Im Neuenheimer Feld 227, 69120 Heidelberg, Germany}
\author{H.~H\"affner}
\affiliation{Department of Physics, University of California, Berkeley, CA 94720, USA}

\begin{abstract}
Quantum statistics have a profound impact on the  properties of systems composed of identical particles. At the most elementary level, Bose and Fermi quantum statistics differ in the exchange phase, either 0 or $\pi$, which the wavefunction acquires when two identical particles are exchanged.
In this Letter, we demonstrate that the exchange phase can be directly probed with a pair of massive particles by physically exchanging their positions. We present two protocols where the particles always remain spatially well separated, thus ensuring that the exchange contribution to their interaction energy is negligible and that the detected signal can only be attributed to the exchange symmetry of the wavefunction. We discuss possible implementations with a pair of trapped atoms or ions.
\end{abstract}

\maketitle 

The symmetrization postulate of quantum mechanics asserts that the wavefunction of a system of identical particles is either completely symmetric or antisymmetric under particle exchange \cite{Messiah:1964}. A plethora of physical phenomena observed in experiments investigating atoms, molecules and solids, as well as the statistical properties of light supports the (anti)symmetrization requirement. While more general quantum statistics \cite{Greenberg:2009} are in principle conceivable, they seem not to be realized by elementary particles in nature \cite{Curceanu:2012}.

The influence of the wavefunction symmetry has been spectacularly demonstrated in few-particle systems with Hong-Ou-Mandel-like interference experiments
\cite{Hong:1987,Bocquillon:2013,Kaufman:2014,Lopes:2015,Islam:2015}, and in many-body systems with ultracold quantum gases \cite{Levin:2012}. Spectroscopic experiments have also tested the symmetrization postulate for massive particles \cite{Kekez:1990,deAngelis:1996,Hilborn:1996,Modugno:1998,CancioPastor:2015} and for photons \cite{DeMille:1999,English:2010} with high precision. Recently, exchange interactions have been applied in engineered quantum systems for entangling pairs of atoms or electrons \cite{Anderlini:2007, Hayes:2007, Kaufman:2015,Veldhorst:2015}.
\begin{figure}[b]
\includegraphics[width=1\columnwidth]{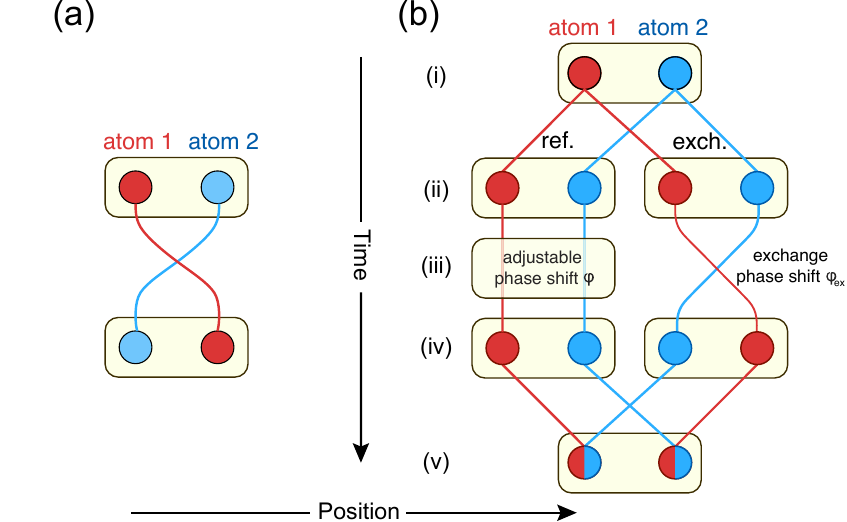}
\caption{\label{fig:SwapParticles} Detection of the wavefunction symmetry in a two-particle interference experiment. (a) Exchanging two identical particles multiplies the wavefunction by a global phase factor $e^{i\varphi_{\rm ex}}=\pm1$, which --- without a reference state --- is not observable. Dynamical and geometrical phases are assumed to vanish. (b) By splitting the wavefunction into a reference path and another path for which the particles' positions are switched, $\varphi_{\rm ex}$ can be detected by correlation measurements after recombining the two paths. The interference signal is controlled by an additional phase $\varphi$, induced by a potential or by the geometry.
}
\end{figure}

At the most elementary level, the wavefunction symmetry manifests itself when two identical particles are exchanged in position [Fig.~\figref{fig:SwapParticles}{a}]: Their state acquires an exchange phase $\varphi_\text{ex}$, which is $0$ for bosons but $\pi$ for fermions.
Exchange of identical particles can naturally occur in molecules where identical, distant nuclei may be interchanged as a result of a rotation \cite{Atkins:2005}.
Prior experiments \cite{Kekez:1990,deAngelis:1996,Hilborn:1996,Modugno:1998,CancioPastor:2015} have exploited this naturally occurring exchange of identical particles to show that only certain rotational states are permitted by the symmetrization postulate.
However, a direct interferometric measurement of the exchange phase $\varphi_\text{ex}$ has never been attempted.
In this Letter, we propose to use the high controllability of trapped atoms or ions for a direct measurement of this phase. To this end, we devise experiments where the two-particle wavefunction  is superposed with the wavefunction of the same particles having swapped positions.
We further request that, if the interferometric sequence is interrupted at any time, the two particles are always found at distant positions.
This condition of vanishing overlap between the two particles ensures that the interference signal depends only on the wavefunction symmetry.

Figure~\figref{fig:SwapParticles}{b} schematically illustrates the general interferometric scheme we envision for detecting  $\varphi_\text{ex}$:
Initially, two identical particles are tightly localized by a confining potential so that their wavefunctions have vanishing overlap. Wavefunction symmetrization plays no role in the description of the initial state since the particles are initially distinguishable by their positions.
Next, by modifying the confining potential, the two-particle wavefunction is split into two parts, a reference state and a state for which the positions of the particles are subsequently swapped.
In the final steps, the two parts of the wavefunction are recombined and two-particle interference is measured.

The scenario sketched in Fig.~\figref{fig:SwapParticles}{b} bears a close resemblance to Hanbury Brown-Twiss \cite{Fano:1961} and Hong-Ou-Mandel \cite{Hong:1987} experiments. However, instead of measuring (anti)bunching of particles as in the majority of these experiments, we will focus on schemes where the two particles are measured at distant sites and interference is detected by correlating the internal or motional states of the atoms.

We present two conceptual ways of realizing the exchange of particles and discuss possible experimental implementations: 
(A) 
A state-dependent potential  transports particles in a way that depends on their internal states.
(B) A state-independent potential confining the atoms is adiabatically transformed. Simultaneously, long-range repulsive interactions such as the Coulomb force between a pair of charged particles  correlate the atom motion in the potential by keeping them apart.

\section{Protocol A: state-dependent transport}
We consider a pair of bosonic or fermionic atoms with two long-lived internal states labelled $|\spinup\rangle$ and $|\spindown\rangle$.
Initially, one atom is prepared at site $S_1$ and the other at site $S_2$, and both of them in the same internal state $|\spinup\rangle$. Their state reads
$a^\dagger_{S_1,\uparrow} a^\dagger_{S_2,\uparrow}\ket{0},$
where $a_{S_i,s}^\dagger$ are the creation operators for the site $S_i$ and pseudo-spin state $\ket{s}$.
We assume that the spatial wavefunctions $\psi_{S_i}(\bm{r}) =\bra{\bm{r},s}a^\dagger_{S_i,s}\ket{0}$ of the two atoms do not overlap.
A $\pi/2$ spin rotation pulse subsequently mixes the internal states and puts the two atoms in a superposition of even- and odd-spin-parity states, $(\ket{\Psi_\text{even}} - \ket{\Psi_\text{odd}})/\sqrt{2}$, defined by
\begin{eqnarray}
\ket{\Psi_\text{even}} &=&2^{-1/2} (a^\dagger_{S_1,\uparrow} a^\dagger_{S_2,\uparrow}+a^\dagger_{S_1,\downarrow} a^\dagger_{S_2,\downarrow})\ket{0},\\
\ket{\Psi_\text{odd}} &=&2^{-1/2} (a^\dagger_{S_1,\uparrow} a^\dagger_{S_2,\downarrow}+a^\dagger_{S_1,\downarrow} a^\dagger_{S_2,\uparrow})\ket{0}.
\end{eqnarray}
Crucially, a physical transport operation conditionally switches the positions of the atoms if they are in, say, $\ket{{\uparrow}}$, while it maintains them at the original location if they are in $\ket{{\downarrow}}$,
\begin{equation}
\label{eq:physicalexchange}
a^\dagger_{S_1,\uparrow}\rightarrow a^\dagger_{S_2,\uparrow}, \hspace{3mm}
a^\dagger_{S_2,\uparrow}\rightarrow a^\dagger_{S_1,\uparrow}, \hspace{3mm}
a^\dagger_{S_i,\downarrow}\rightarrow e^{i\varphi/2} a^\dagger_{S_i,\downarrow},
\end{equation}
where we also allow for a precisely adjustable dynamical phase $\varphi$ acquired during the process.
To ensure vanishing exchange interactions, the exchange process must be realized such that  $\psi_{S_i}(\bm{r};t)$ remain disjoint for all times $t$, i.e., $\psi_{S_1}(\bm{r};t)\psi_{S_2}(\bm{r};t) = 0$.

The evolution of $\ket{\Psi_\text{even}}$ under the transformation in Eq.~(\ref{eq:physicalexchange}) realizes the situation sketched in Fig.~\figref{fig:SwapParticles}{b}.
The correspondence is apparent once the different terms are reordered according to the commutation rules $a^\dagger_{S_1,s}a^\dagger_{S_2,s}=e^{i\varphi_\text{ex}}a^\dagger_{S_2,s}a^\dagger_{S_1,s}$, yielding
\begin{equation}
	\label{eq:evolution}
\ket{\Psi_\text{even}} \rightarrow \frac{1}{\sqrt{2}} (e^{i\varphi_\text{ex}}a^\dagger_{S_1,\uparrow} a^\dagger_{S_2,\uparrow}+e^{i\varphi}a^\dagger_{S_1,\downarrow} a^\dagger_{S_2,\downarrow})\ket{0}.
\end{equation}
Thus, the exchange phase $\varphi_{\rm ex}$ now appears in the description of the internal state as a relative phase, which can be detected by correlating local measurements of the particles' internal state \cite{Sackett:2000}: after applying a second $\pi/2$ spin rotation pulse, the expectation value of the spin parity operator $\Pi$ \cite{ParityOperator} yields $\langle{\Pi}\rangle=\cos(\varphi-\varphi_\text{ex})$. 
Recording $\Pi$ for different values of $\varphi$ allows one to measure $\varphi_\text{ex}$.
The evolution of $\ket{\Psi_\text{odd}}$ is different, though, and leads to a state with two atoms in the same location, where the exchange phase (as well as the dynamical phase) has no influence on spin correlations between the two particles.
If not discarded through post-selection, these events would halve the visibility of the parity signal.
If  state $\ket{\Psi_\text{even}}$ is directly prepared using an entangling scheme for distant particles, see Refs.~\cite{Wilk:2010,Maller:2015}, full visibility of the spin-parity fringe can be ideally obtained without post-selection.

\section{Implementation with a pair of neutral atoms in an optical lattice}
Protocol A can be realized using a pair of distant neutral atoms that are transported in spin-dependent optical lattices \cite{Deutsch:1998,Mandel:2003,Steffen:2012,Robens:2017} ; other forms of state-dependent transport with microwave-dressed potentials in atom chips \cite{Boehi:2009} or with spin-dependent optical tweezers are also conceivable.

\begin{figure}[t]
\includegraphics[width=\linewidth]{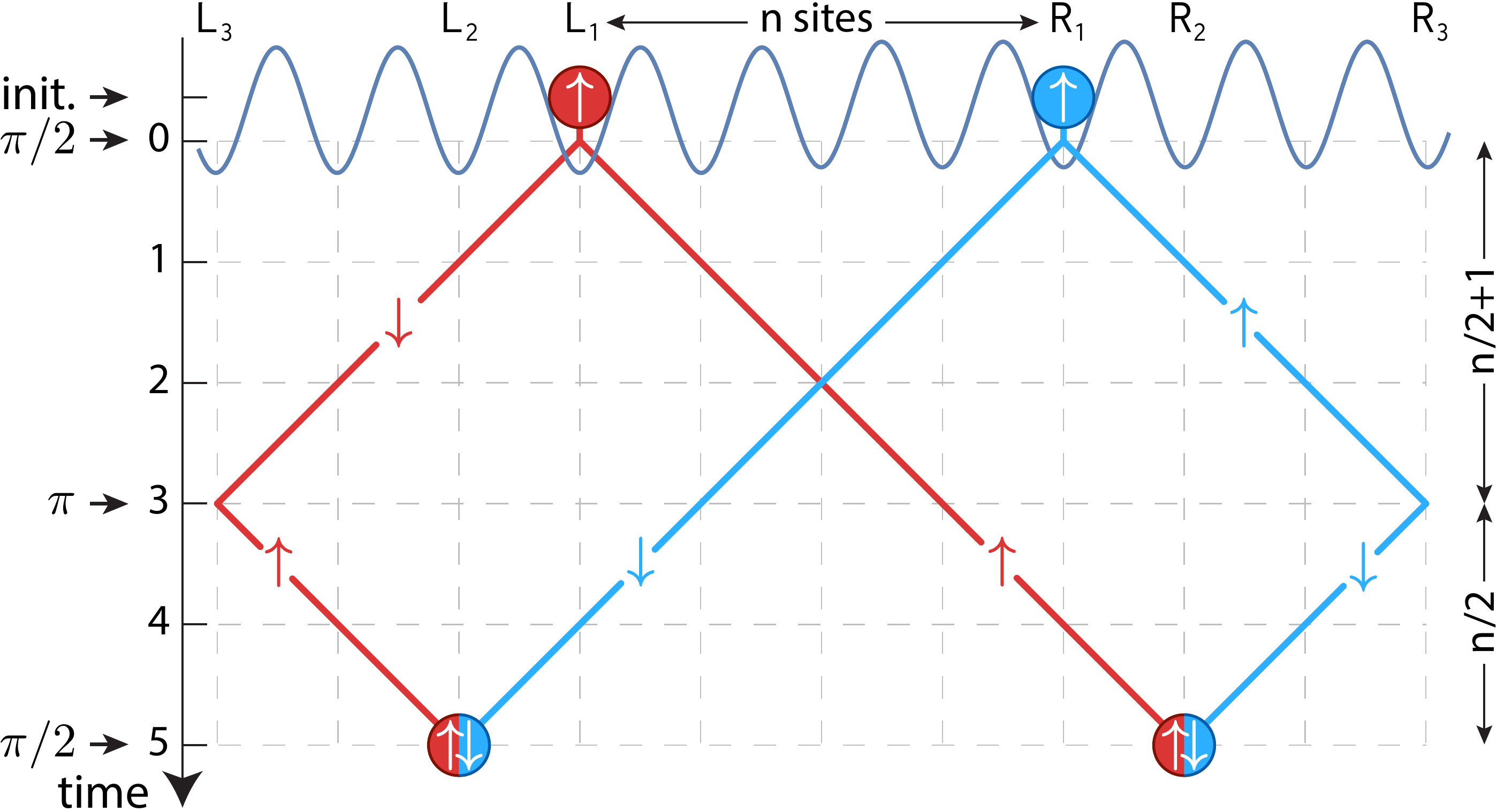}
\caption{\label{fig:NeutralAtomFig} Two-atom Ramsey interferometer sequence probing quantum statistics with two distant neutral atoms. A spin parity measurement produces a two-atom Ramsey-like fringe, whose phase depends on  $\varphi_\text{ex}$. To recombine the atoms, a position-dependent $\pi$ pulse \cite{Weitenberg11,Karski:2010, Xia:2015} is applied to the outermost sites, $L_3$, $R_3$, see Ref.~\cite{SupplementaryMaterial}. The arrows indicate the spin state for the different paths, $n$ denotes the initial separation, and time is expressed in units of shift operations. A two-dimensional variant is presented in the Supplemental Material \cite{SupplementaryMaterial}, which ensures that the two atoms stay always far apart.
} 
\end{figure}

We propose a two-particle Ramsey interferometer as is shown in Fig.~\ref{fig:NeutralAtomFig}, which, instead of probing first-order coherence, detects second-order coherence revealing $\varphi_\text{ex}$: A pair of atoms is initially prepared in well-separated lattice sites, denoted $L_1$ and $R_1$, with their pseudospin states in $|\spinup\rangle|\spinup\rangle$.
The lattice depth is chosen sufficiently high to suppress tunneling to neighboring sites \cite{Steffen:2012}.
Importantly, both atoms must be cooled to the lowest vibrational state of their respective lattice potential well \cite{Kaufman:2012,Li:2012,Robens:2017}, in order to make them indistinguishable in the motional degree of freedom, see Supplemental Material \cite{SupplementaryMaterial}.
The first $\pi/2$ Ramsey pulse puts both atoms in a superposition of $\ket{{\uparrow}}$ and $\ket{{\downarrow}}$ states.
Subsequently, each atom is split in space and transported conditioned upon its pseudospin state \cite{Robens:2017} to both end sites $L_2$ and $R_2$.
Each shift operation can be performed fast on the time scale of $\SI{10}{\micro\second}$ per lattice site \cite{Steffen:2012}. In particular, polarization-synthesized optical lattices \cite{Robens:2017} allow one to  state-dependently transport atoms in a single operation over few tens of lattice sites, while at the end leaving the atoms in the lowest vibrational state \cite{Robens:2016c}.
Finally, the second $\pi/2$ Ramsey pulse erases the information about which way the atoms travelled to reach the end sites.
Focusing our attention on atoms detected at distant sites \cite{Alberti:2016,Robens:2016}, local spin measurements yield an equal probability to find $\ket{{\uparrow}}$ or $\ket{{\downarrow}}$, meaning that each atom probed individually is found in a statistical mixture of both spin states.
However, a parity measurement of the spin state \cite{ParityOperator} yields nontrivial correlations, showing for example perfect spin alignment for bosonic and antialignment for fermionic atoms.
An interference fringe can be recorded by precisely adjusting the phase difference $\varphi$ between the outermost and innermost paths, for example, by controlling the relative phase of the position-dependent \cite{SupplementaryMaterial} pulse acting at sites $L_3$ and $R_3$. 
With $\SI{90}{\percent}$ of atoms prepared in the lowest vibrational state, we expect a visibility of the spin parity signal of $\approx \SI{80}{\percent}$ \cite{SupplementaryMaterial}.
Note that while the Ramsey scheme in Fig.~\ref{fig:NeutralAtomFig} preserves the connectedness of the abstract protocol sketched in Fig.~\figref{fig:SwapParticles}{b}, it is designed to be robust against dephasing mechanisms. Stochastic dynamical phases caused by fluctuating magnetic fields, magnetic field gradients, and state-dependent transport operations cancel out owing to time and space refocusing \cite{SupplementaryMaterial}.

Remarkably, nontrivial correlations are predicted in the proposed scheme even though the two particles have never met nor interacted with each other. These correlations are purely quantum and, as such, incompatible with a macro-realistic worldview \cite{Bassi:2013} where atoms travel either the outermost or the innermost paths. Correlations from accidental interactions between the two atoms at the intersection point in the center can be made vanishingly small by increasing the transport velocity and by softening the transverse confinement; in a two-dimensional scheme using two-dimensional spin-dependent optical lattices \cite{Groh:2016}, interactions are completely avoided, see Ref.~\cite{SupplementaryMaterial}.

Conceptually, the closest analog to this scheme is the Franson interferometer \cite{Franson:1989} suggested to test local hidden-variable theories with two photons independently emitted at consecutive times. However, here the massive particles are ``emitted'' (namely, transported) simultaneously. It also shares a resemblance with Fano's interpretation \cite{Fano:1961} of Hanbury Brown and Twiss' experiment, although here we detect spin correlations instead of (anti)bunching of particles.
As a potential application, the proposed interference scheme would allow one to test nonlocal correlations \cite{Clauser:1969,Clauser:1978} between macroscopically distant atoms \cite{Kaufman:2015}.
We expect that entangling atoms separated by macroscopic distances on the scale of few thousands of lattice sites should be doable with currently available technology \cite{Robens:2017}.

Suitable atomic species for such experiments are discussed in detail in Ref.~\cite{SupplementaryMaterial}: Rb \cite{Mandel:2003} and Cs \cite{Steffen:2012} for bosons, and alkali-earth-like atoms \cite{Daley:2011} for fermions. Moreover, aluminum is an attractive atomic species for a direct comparison of the exchange phase of fermionic ($^{26\!}$Al) and bosonic ($^{27\!}$Al) isotopes with the same experimental setup.

\begin{SCfigure*}
\includegraphics[width=0.6\textwidth]{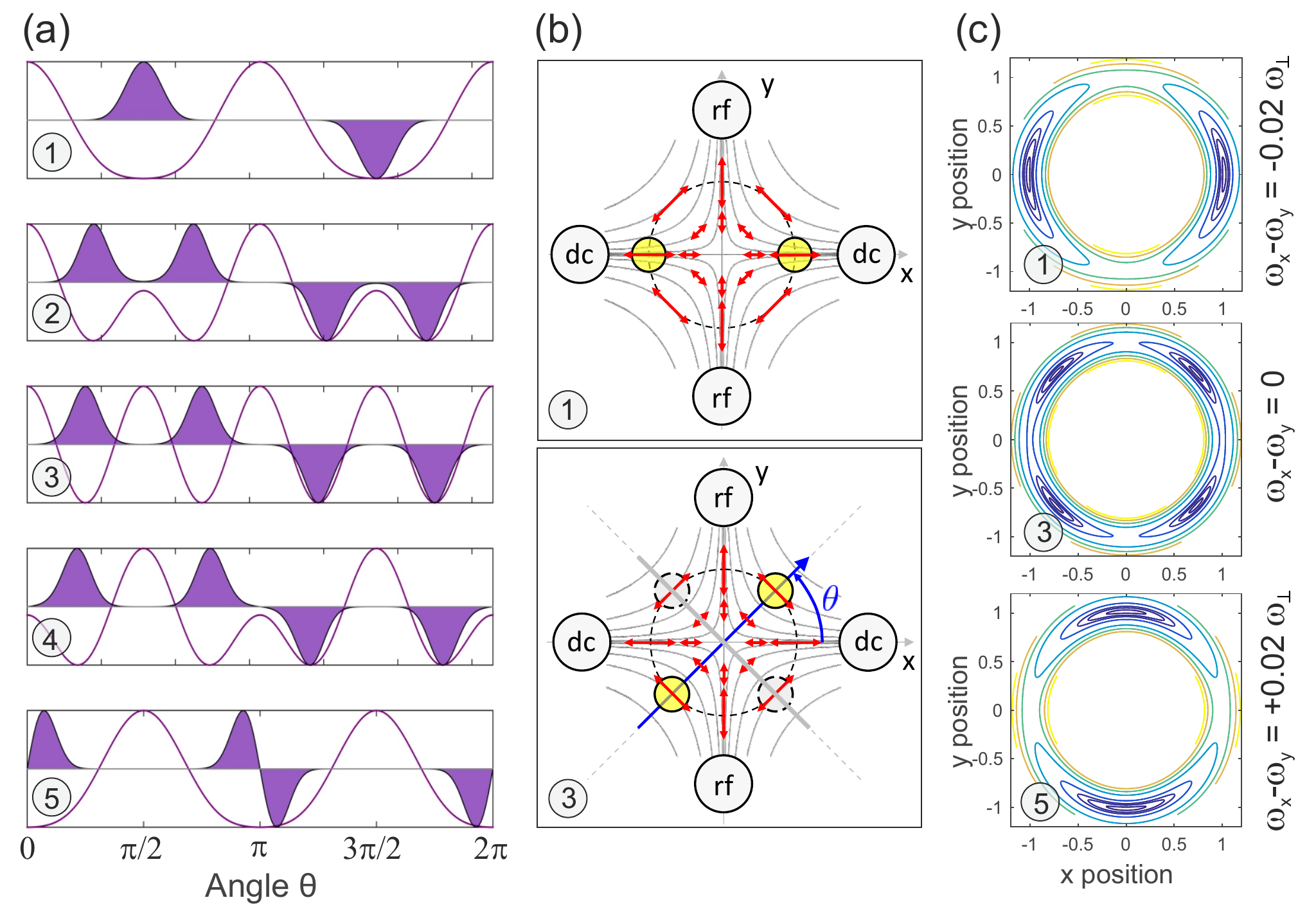}
\caption{\label{fig:Figure3.pdf} Trapped-ion protocol. (a) The two-ion wavefunction is treated as a one-dimensional quantum rotor (shown for fermions in the same internal state). An adiabatic transformation splits a single-well (1) into a double well potential (3) that is subsequently merged again into a single well, but at a different position (5). In the case of fermions (bosons), the final state (5) has opposite (same) parity as compared to that of the initial state (1).  (b) In the radial rf-quadrupolar potential of a linear trap, a two-ion rotor (yellow spheres) can be aligned with the x-axis by dc-voltages reducing the confinement along the x-axis (top). For zero dc-voltage, the radial symmetry of the confining potential is broken by the orientation of the micromotion (red arrows) with respect to the rotor axis \cite{SupplementaryMaterial}. The rotor will align under an angle $\theta=\pm \pi/4$ with respect to the x-axis (bottom). (c) Contour lines of the time-averaged Coulomb and trapping potential as a function of the relative position vector ${\bm{r}}$ for three different potentials.}
\end{SCfigure*}

\section{Protocol B: long-range interactions}
In the presence of a confining potential, long-range repulsive interactions turn two particles into a molecule-like quantum rotor. We  assume  a  potential  that  is  strongly  confining in one dimension, effectively freezing out the rotor’s motion in this direction, and that has a single minimum in the orthogonal plane. As in the case of homonuclear diatomic molecules, the symmetry of the spin state here also controls the symmetry of the spatial wavefunction $\Psi(\theta)$ of the rotor \cite{Deep:2007,Fleischer:2007} with orientation angle $\theta$ in the weakly confining plane. For clarity, we focus  on the case of fermionic particles.
If the particles are prepared in, for example, $\ket{\spindown}\ket{\spindown}$, the rotor's wavefunction must be antisymmetric, $\Psi(\theta)=-\Psi(\theta+\pi)$, as sketched in Fig.~\figref{fig:Figure3.pdf}{a}. Apart from that, the spin state plays no role in this protocol
in contrast to the previous one. Although the wavefunction can be completely specified by limiting the angle to a range of $0\le\theta<\pi$, it is convenient to represent it over the full range $0\le\theta<2\pi$.

We assume that the rotor is initially prepared in the ground state of the potential $V(\theta,t=0)=V_0\cos^2\theta$.  If the initial potential well located, e.g., at $\theta=\pi/2$ is adiabatically split into a double well, the Gaussian wave packet of the ground state will be transformed into an even superposition of wave packets. By slowly separating the two minima of the double well (and of the double well at $\theta=3\pi/2$ as well), the wave packets originating from opposite sides of the ring will eventually meet and merge into a wave packet with uneven parity. The final potential again consists of a single well, but now located at $\theta=\pi$ (or $\theta=2\pi$). Importantly, for spatial wavefunctions that are antisymmetric under particle exchange, the adiabatic transport maps the even states of the initial potential onto odd states of the final potential and vice versa, whereas for spatially symmetric wavefunctions the state's parity is preserved. Because of this property,
bosonic and fermionic atoms can be distinguished by measuring whether the parity of the motional state has changed at the end of the adiabatic transport. The analysis of the rotor's angular motional state is equivalent to a correlation measurement of local modes of motion of the two atoms.

Such a rigid ion-rotor behaves similarly to homonuclear diatomic molecules, where techniques such as pendular state spectroscopy or rotational coherence spectroscopy reveal the effect of the exchange symmetry on allowed rotational states. In contrast, in the experiment proposed here, complete control over the rotor enables exchanging the particles without rotating their electronic wavefunction \cite{Atkins:2005} and, most importantly, a direct measurement of the exchange phase.

\section{Implementation with a pair of trapped ions}
For the realization of protocol B, we consider a linear radio-frequency (rf) trap confining the ions in a harmonic potential with oscillation frequencies $\omega_x$, $\omega_y$ in the radial directions and $\omega_z$ in the axial direction. The difference between the radial oscillation frequencies can be controlled by a static voltage $U_\text{dc}$. A pair of laser-cooled ions forms a crystal in the radial plane if $\omega_z>\omega_{x}, \omega_{y}$. At the ions' equilibrium positions, the trapping force is balanced by the ions' mutual Coulomb repulsion. 

Because of the harmonic confinement, the ion dynamics separates into the center-of-mass motion and the relative motion ${\bm{r}}={\bm{r}}_1-{\bm{r}}_2$. 
The latter is governed by the Hamiltonian
\[
H_r = -\frac{\hbar^2}{2\mu}\nabla_r^2+\frac{\mu}{2}(\omega_x^2r_x^2+\omega_y^2r_y^2) + V_{\rm coul}({\bm{r}}),
\]
where $\mu$ is the ions' reduced mass and $r_x$, $r_y$ are the transverse components of ${\bm{r}}$. Due to the micromotion of the ions in the radial plane, one has to time-average the Coulomb energy over one period of the rf-driving field, leading to a modified Coulomb potential 
\[
V_{\rm coul}(\vec{r})= \frac{e^2}{4\pi\epsilon_0|\vec{\bm{r}}|}\left(1+\frac{3}{16}q^2\cos^2(2\theta)\right)
\]
where $q$ is the trap's $q$-parameter \cite{Raizen:1992a} and $\theta$ denotes the orientation of the crystal in the radial plane (see Fig.~\figref{fig:Figure3.pdf}{b} and \cite{SupplementaryMaterial}). For the case where $\omega_x=\omega_y$, the asymmetry of the micromotion lifts the rotational symmetry of the potential and leads to two equivalent sets of equilibrium orientations of the crystal under $\theta=\pm\pi/4$ \cite{Moore:1994,Hoffnagle:1995}.

This effect opens up the prospect of implementing protocol $B$ with a pair of ions by ramping $U_\text{dc}$ from positive to negative voltages. The relative motion is described by two normal modes, which we assume to be cooled to the ground state.
If initially $\Delta\omega=\omega_y-\omega_x>0$, the
ion crystal is aligned with the $x$-axis [Fig.~\figref{fig:Figure3.pdf}{b}, top]. Lowering $\Delta\omega$ by reducing $U_\text{dc}$
softens the normal mode perpendicular to the  crystal's axis (the rocking mode) while hardly affecting the other mode.
At the critical value $\Delta\omega_\text{crit}=\frac{3}{4}q^2\omega_\perp$, with $\omega_\perp=\sqrt{(\omega_x^2+\omega_y^2)/2}$, the rocking mode potential becomes quartic and subsequently splits into a double well. The wells separate and move to $\theta=\pm\pi/4$ when $U_\text{dc}$ becomes zero [Fig.~\ref{fig:Figure3.pdf}(b), bottom, and ~\ref{fig:Figure3.pdf}(c)]. At this point, the ion rotor is in a coherent superposition of two perpendicular orientations. Ramping $U_\text{dc}$ to negative values combines a different pair of wells which will finally merge, resulting in an ion rotor oriented along the $y$-axis. In this way, the two possible paths of rotating the ion rotor clockwise or counter-clockwise interfere, and a measurement by sideband spectroscopy of the motional state \cite{Diedrich:1989} of the rocking mode reveals the bosonic or fermionic character of the ions. For further information, see Ref.~\cite{SupplementaryMaterial}.

The quantum coherence of the process can be checked by initially preparing the internal state of the ions in a Bell state $ (|\spinup\rangle|\spindown\rangle+e^{i\phi}|\spindown\rangle|\spinup\rangle)/\sqrt{2}$. The phase $\phi$ controls the symmetry of the spatial wavefunction \cite{Zeilinger:1998}, which for the special case $\phi=0$ ($\pi$) is antisymmetric  (symmetric). As a consequence, this phase determines whether the protocol maps the rocking mode's state onto the ground or first excited state.

The experiment could be carried out with ion species like the fermionic $^{40}$Ca$^+$ or the bosonic $^{43}$Ca$^+$, for which ground state cooling and Bell state generation are routinely done  \cite{Schindler:2013,Ballance:2016}. A numerical simulation of the time-dependent Schr\"odinger equation suggests that an adiabatic transfer is achievable in less than 2~ms
 \cite{SupplementaryMaterial}, much shorter than the time scale on which heating of the relative ion motion occurs.

In the absence of imperfections, this protocol constitutes an interferometer with completely symmetric arms. Therefore, it should be immune against dynamical phases. A non-zero magnetic flux through the circle on which the ions move, however, would lift the symmetry and give rise to a small, but measurable geometric Aharonov-Bohm phase \cite{Noguchi:2014}. 

An experimental challenge is to suppress stray electric field gradients, which, by breaking the symmetry of the confining potential, would cause dynamical phases or even compromise the process of splitting the minimum of the potential into two.  After compensation of such fields, it should be possible to independently measure the remaining dynamical phases (see Ref.~\cite{SupplementaryMaterial}) in order to extract $\varphi_\text{ex}$ from the measured signal.

The proposed protocol shows that quantum statistics can become important for trapped ions \cite{Yin:1995} in experimentally accessible parameter regimes. A quantum gate entangling the pair of ions based solely on particle exchange could be realized by first carrying out the protocol and then running it backwards again after a suitable waiting time. Since triplet and singlet states have different symmetry, and therefore are transiently mapped to different motional states, they pick up different phase factors. In this way, a $\sqrt{\mbox{SWAP}}$ gate could be realized as used for solid-state quantum computing based on exchange interactions \cite{Loss:1998, Kane:1998, Veldhorst:2015} (and for linear-optical quantum information processing \cite{Cernoch:2008}). The protocol could even be applied to a pair of molecular ions. In addition, it could lead to ion-based quantum sensors complementing single-particle interferometry schemes based on structural phase transitions \cite{Retzker:2008, Shimshoni:2011}.

\section{Conclusions}
The proposed experiments show that the exchange phase can be precisely measured with massive particles. By ensuring that the particles' wavefunctions have vanishing overlap, a situation not encountered so far in Hong-Ou-Mandel-like experiments \cite{Hong:1987, Heeres:2013,Bocquillon:2013, Kaufman:2014, Lopes:2015, Islam:2015,Toyoda:2015}, these experiments would demonstrate the effect of exchanging two identical particles at the most elementary level. Moreover, the two protocols open novel perspectives for entanglement generation and sensing applications based on a pair of identical particles.

\let\section\origsection
\let\addcontentslineorig\addcontentsline

\renewcommand{\addcontentsline}[3]{}

\begin{acknowledgments}   
C.\,R.\ acknowledges helpful discussions with P.~Zoller on analogies between the trapped-ion proposal and molecular physics. The authors wish to thank W.~Alt, M.~Baranov and the anonymous referees for insightful discussions and helpful comments. A.\,A.\ and D.\,M.\ acknowledge financial support from the Deutsche Forschungsgemeinschaft (SFB/ TR 185 OSCAR) and the ERC advanced grant DQSIM (Project ID 291401). P.\,H.\ acknowledges support from the Austrian Science Fund (FWF) through the SFB FoQuS (Project No.\ F4016-N23), and from the European Commision through the ERC synergy grant UQUAM and the ERC advanced grant EntangleGen (Project ID 694561).

\vspace{1.5mm}C.\,R. and A.\,A. contributed equally to this work.
\end{acknowledgments}


\section*{Supplemental Material}

\makeatletter
\renewcommand*{\thesection}{S\arabic{section}}
\renewcommand*{\thesubsection}{\Alph{subsection}}
\renewcommand*{\p@subsection}{\thesection.}
\renewcommand*{\thesubsubsection}{\arabic{subsubsection}}
\renewcommand*{\p@subsubsection}{\thesubsection.}
\makeatother
\renewcommand{\thefigure}{S\arabic{figure}}
\renewcommand{\thetable}{S\arabic{table}}
\renewcommand{\theequation}{S\arabic{equation}}
\setcounter{equation}{0}
\setcounter{figure}{0}

\makeatletter
\renewcommand\paragraph[1]{\@startsection{paragraph}{4}{\parindent}%
{\z@}%
{-\lineskip}%
{\itshape\normalsize}*{{#1}.\textemdash}}%
\makeatother
\renewcommand{\theparagraph}{}

The following sections provide additional information about the protocols for a pair of distant, trapped neutral atoms or ions. The contents are organized as follows:\vspace{-1.4cm}

\def\tocname{}
\tableofcontents

\let\addcontentsline\addcontentslineorig

\vspace*{-2mm}\section{Protocol for neutral atoms}

\subsection{Two-atom Ramsey interferometer}
\label{sec:RamseyFringe}
We provide a detailed description of the two-atom Ramsey sequence producing the spin parity signal $\langle \Pi \rangle$, which carries information about the exchange phase $\varphi_\text{ex}$.

\paragraph{State preparation}We assume that the two atoms are initially prepared with internal state $\ket{{\uparrow}}$, and situated at a distance of $n$ lattice sites apart \cite{Robens:2017}.
Their state is represented by the product state
\begin{equation}
	\label{eq:productstate}
	a^\dagger_{L_1,\uparrow} a^\dagger_{R_1,\uparrow}\ket{0},
\end{equation}
where \encapsulateMath{$a^\dagger_{L_i,s}$} and \encapsulateMath{$a^\dagger_{R_i,s}$} are the creation operators at sites $L_i$ and $R_i$ indicated in Fig.~2 of the main text, with internal state $\ket{{s}}$.

The first $\pi/2$ pulse of the Ramsey sequence superposes the two pseudo-spin states $\ket{{\uparrow}}$ and $\ket{{\downarrow}}$, realizing the unitary transformation 
\begin{equation}
	\label{eq:firstpulse}
	U_{\pi/2}(x) =\frac{1}{\sqrt{2}}\left(\hspace{-3pt}\begin{array}{cc} 1& e^{-i\hspace{1pt}\varphi_x}\\ -e^{i\hspace{1pt}\varphi_x} & 1\end{array}\hspace{-3pt}\right)\hspace{-1pt},\hspace{2pt} \text{with $x=L_1,R_1$,}
\end{equation}
where the unitary matrix is specified in the rotating frame in which the two pseudo-spin states are degenerate in energy,  and $\varphi_x$ are two position-dependent phases.
We assume that position-resolved pulses, as detailed in Sec.~\ref{sec:posdeppulses}, allow one to individually control these phases for each site, $L_1$ and $R_1$, independently.
The $\pi/2$ pulse transforms the initial product state, see Eq.~(\ref{eq:productstate}), into
\begin{multline}
	\label{eq:afterpipulse}
		\hspace{-4mm} \left[U_{\pi/2}(L_1) a^\dagger_{L_1,\uparrow} U_{\pi/2}^\dagger(L_1)\right]\!\left[U_{\pi/2}(R_1) a^\dagger_{R_1,\uparrow}U_{\pi/2}^\dagger(R_1)\right] \ket{0} =\\(\ket{\Psi_\text{even}} - \ket{\Psi_\text{odd}})/\sqrt{2},
\end{multline}
which is the superposition of two nonseparable states with even and odd spin parity,
\begin{eqnarray}
	\label{eq:psieveninit}
	\ket{\Psi_\text{even}} &=&\frac{a^\dagger_{L_1,\uparrow} a^\dagger_{R_1,\uparrow}+e^{i\,(\varphi_{L_1}+\varphi_{R_1})}a^\dagger_{L_1,\downarrow} a^\dagger_{R_1,\downarrow}}{\sqrt{2}}\ket{0},\\
	\label{eq:psioddinit}
	\ket{\Psi_\text{odd}} &=&\frac{e^{i\varphi_{R_1}}a^\dagger_{L_1,\uparrow} a^\dagger_{R_1,\downarrow}+e^{i\varphi_{L_1}}a^\dagger_{L_1,\downarrow} a^\dagger_{R_1,\uparrow}}{\sqrt{2}}\ket{0}.
\end{eqnarray}
The square brackets in Eq.~(\ref{eq:afterpipulse}) are used to emphasize that the total state is still separable since, in fact, product states of noninteracting particles remain separable under unitary evolution.

The common phase $(\varphi_{L_1}\hspace{1pt}{+}\hspace{1pt}\varphi_{R_1})/2$ only affects first-order coherence, and plays no role in the proposed two-atom Ramsey interferometer, which exclusively probes second-order coherence.
In contrast, the relative phase $\Delta \varphi_1 = \varphi_{L_1}\hspace{1pt}{-}\hspace{1pt}\varphi_{R_1}$, which originates from a local transformation of the internal states, does influence the two-atom interference signal since it determines how the two atoms behave under exchange of positions:
As this phase is varied, the state in Eq.~(\ref{eq:psioddinit}) periodically changes between spin triplet $\ket{\Psi_\text{T}}$ and spin singlet $\ket{\Psi_\text{S}}$ states:
\begin{eqnarray}
	\ket{\Psi_\text{odd}} &=& \frac{(e^{i\varphi_R}+e^{i\varphi_L})\ket{{\Psi_\text{T}}}+(e^{i\varphi_R}-e^{i\varphi_L})\ket{{\Psi_\text{S}}}}{\sqrt{2}},
\end{eqnarray}
with
\begin{eqnarray}
	\label{eq:triplet}
	\ket{\Psi_\text{T}} &=&\frac{a^\dagger_{L_1,\uparrow} a^\dagger_{R_1,\downarrow}+a^\dagger_{L_1,\downarrow} a^\dagger_{R_1,\uparrow}}{\sqrt{2}}\ket{0},\hspace{1.65cm}\\
	\label{eq:singlet}
	\ket{\Psi_\text{S}} &=&\frac{a^\dagger_{L_1,\uparrow} a^\dagger_{R_1,\downarrow}-a^\dagger_{L_1,\downarrow} a^\dagger_{R_1,\uparrow}}{\sqrt{2}}\ket{0}.
\end{eqnarray}
Notably, under exchange of positions, $L_1\leftrightarrow R_1$, the spin singlet state acquires a $\pi$ phase, which adds on top of the exchange phase $\varphi_\text{ex}$.
This means that for $\Delta \varphi_1=\pi$, when $\ket{{\Psi}_\text{odd}}$ is a singlet, the two atoms behaves as if they switched to opposite quantum statistics \cite{Zeilinger:1998}.

It is important that the evolution of the two states in Eqs.~(\ref{eq:psieveninit}) and (\ref{eq:psioddinit}) can be unambiguously discriminated, allowing one to single out through post-selection the events associated with $\ket{{\Psi_\text{odd}}}$, which yield a two-atom correlation signal varying with $\Delta\varphi_1$.
As an alternative to post-selection, one can start the Ramsey sequence directly with the nonseparable state $\ket{{\Psi_\text{odd}}}$, which can be prepared with entangling schemes for distant particles, see Refs.~\cite{Wilk:2010,Maller:2015}; in these schemes, long-range interactions, which are used to entangle the two particles, are insensitive to the fermionic or bosonic statistics.

\paragraph{Exchange of positions} There is a host of spin-dependent transport sequences that allow the swap of the two atoms and that are also robust against common decoherence mechanisms, see Sec.~\ref{sec:decoherence}.
For simplicity, we restrict ourselves here to the sequence sketched in Fig.~2 of the main manuscript, which comprises $n/2+1$ shift operations,  position-dependent $\pi$ pulses acting at the outermost sites, and further $n/2-1$ shift operations.
The different number of shift operations avoids that atoms in the innermost paths occupy the same lattice site for the duration of the $\pi$ pulse. This minimizes the time atoms spend near each other, where they can interact through onsite collisions. By transporting the atoms on paths in two spatial dimensions, see Sec.~\ref{sec:twodim}, interactions are fully avoided since atoms never meet.

To recombine the atoms, the position-dependent $\pi$ pulse, see Sec.~\ref{sec:posdeppulses}, is applied only to the outermost sites $L_3$, $R_3$, realizing the unitary transformation
\begin{equation}
	U_{\pi}(x) =\left(\begin{array}{cc} 0& e^{i\hspace{1pt}\varphi_x}\\ -e^{-i\hspace{1pt}\varphi_x} & 0\end{array}\right),\hspace{2mm} \text{with $x=L_3,R_3$.}
\end{equation}
For convenience of notation, the phase $\varphi_x$ in $U_\pi(x)$ is expressed with opposite sign convention with respect to that used in Eq.~(\ref{eq:firstpulse}) for $U_{\pi/2}(x)$. 

The entire exchange process, including $n$ spin-dependent shift operations and $\pi$ pulse, results in the unitary transformation
\begin{align}
	\label{eq:transport1}
a^\dagger_{L_1,\uparrow}&\rightarrow a^\dagger_{R_2,\uparrow}\,,& 
a^\dagger_{L_1,\downarrow}&\rightarrow e^{i\hspace{1pt}\varphi_{L_3}} a^\dagger_{L_2,\uparrow}\,,\\
a^\dagger_{R_1,\uparrow}&\rightarrow  -e^{-i\hspace{1pt}\varphi_{R_3}} a^\dagger_{R_2,\downarrow}\,,&
a^\dagger_{R_1,\downarrow}&\rightarrow a^\dagger_{L_2,\downarrow}\,.\nonumber
\end{align}
Dynamical phases acquired during transport are shown in Sec.~\ref{sec:decoherence} to have no effect on the spin parity signal; for simplicity, we set them equal to zero in Eq.~(\ref{eq:transport1}).
The transformation in Eq.~(\ref{eq:transport1}) yields the evolution
\begin{eqnarray}
	\label{eq:transportevolutionsamesite}
	\hspace{-0.5cm}\ket{\Psi_\text{even}} &\rightarrow&\nonumber\\
	&&\hspace{-1.5cm}
		\frac{a^\dagger_{R_2,\uparrow} a^\dagger_{R_2,\downarrow}-e^{i(\varphi_{L_1}+\varphi_{R_1}+\varphi_{L_3}+\varphi_{R_3})}a^\dagger_{L_2,\uparrow}  a^\dagger_{L_2,\downarrow}}{\sqrt{2}}\ket{0},\\
	\label{eq:separatesites}
	\hspace{-24pt}\ket{\Psi_\text{odd}} &\rightarrow&\nonumber\\
	&&\hspace{-1.5cm}
		\frac{ e^{i(\varphi_\text{ex}-\Delta\varphi_1-\Delta\varphi_3)}a^\dagger_{L_2,\downarrow} a^\dagger_{R_2,\uparrow}-a^\dagger_{L_2,\uparrow}a^\dagger_{R_2,\downarrow}}{\sqrt{2}}\ket{0},
\end{eqnarray}
where $\Delta\varphi_3 = \varphi_{L_3}-\varphi_{R_3}$ is equal to the relative phase of the two $\pi$ pulses at sites $L_3$ and $R_3$.
For position-independent microwave pulses (i.e., pulses acting uniformly at both sites), such a relative phase is very small, of the order of $2\pi\,\num{e-5}$, due to the centimeter-long wavelength, and can be precisely accounted for.
For position-dependent pulses, crass-talk-induced phase corrections of the order of $2\pi\,\num{e-2}$ are computed in Sec.~\ref{sec:posdeppulses}.
Unimportant global phases have been discarded in the expression of the two states.

As anticipated, Eqs.~(\ref{eq:transportevolutionsamesite}) and (\ref{eq:separatesites}) show that the two states evolve differently: two atoms in $\ket{{\psi_\text{odd}}}$ travel to two distant lattice sites ($L_2$ and $R_2$), whereas two atoms in $\ket{{\psi_\text{even}}}$ end up in the same site (either $L_2$ or $R_2$).
The exchange phase $\varphi_\text{ex}$ results from reordering the creation operators in Eq.~(\ref{eq:separatesites}).
Note also that by varying either $\Delta \varphi_1$ or  $\Delta \varphi_2$, the state in Eq.~(\ref{eq:separatesites}) periodically changes between triplet and singlet states akin to those defined in Eqs.~(\ref{eq:triplet}) and (\ref{eq:singlet}).

\paragraph{Spin parity measurement}
Fluorescence imaging techniques \cite{Alberti:2016} allow one to discriminate with high accuracy events with two atoms in the same site, see Eq.~(\ref{eq:transportevolutionsamesite}), from those with two atoms detected in distant sites, see Eq.~(\ref{eq:separatesites}).
Only events with distant atoms are post-selected to measure spin correlations.
The spin state of the two atoms at sites $L_2$ and $R_2$ can be probed with a nondestructive spin measurement technique \cite{Robens:2016}, which employs spin-dependent optical lattices to perform an optical Stern-Gerlach spin detection.

Both states in Eqs.~(\ref{eq:transportevolutionsamesite}) and (\ref{eq:separatesites}) have odd spin parity, and their spin state directly reveals which paths \textemdash either the outermost or the innermost ones \textemdash the two atoms have traveled. Thus, to produce second-order coherence, a final $\pi/2$ pulse is applied, which erases the which-way information. The pulse realizes the unitary transformation $U_{\pi/2}(x)$, see Eq.~(\ref{eq:firstpulse}), with $x=L_2,R_2$.

The state with atoms at distant sites, see Eq.~(\ref{eq:separatesites}), evolves as a result of the $\pi/2$ pulse in a superposition of states with odd and even spin parity. For our purpose, it is sufficient to consider the projections onto different parity states:
\begin{eqnarray}
	\label{eq:finalodd}
	\ket{\Psi_\text{odd}} \xrightarrow[\text{odd spin parity}]{\text{projection onto}}&\nonumber\\
	&&\hspace{-4cm}
		\cos\hspace{-2pt}\left(\hspace{-2pt}\frac{\varphi_\text{ex}{-}\varphi}{2}\hspace{-2pt}\right)\hspace{-2pt}\frac{e^{i\varphi_{R_2}}a^\dagger_{L_2,\uparrow} a^\dagger_{R_2,\downarrow}-e^{i\varphi_{L_2}}a^\dagger_{L_2,\downarrow} a^\dagger_{R_2,\uparrow}}{\sqrt{2}}\ket{0},\\
	\label{eq:finaleven}
	\ket{\Psi_\text{odd}} \xrightarrow[\text{even spin parity}]{\text{projection onto}}&\nonumber\\
	&&\hspace{-4cm}
		\sin\hspace{-2pt}\left(\hspace{-2pt}\frac{\varphi_\text{ex}{-}\varphi}{2}\hspace{-2pt}\right)\hspace{-2pt}\frac{a^\dagger_{L_2,\uparrow} a^\dagger_{R_2,\uparrow}-e^{i\,(\varphi_{L_2}+\varphi_{R_2})}a^\dagger_{L_2,\downarrow} a^\dagger_{R_2,\downarrow}}{\sqrt{2}}\ket{0},
\end{eqnarray}
where $\varphi = \Delta\varphi_1 + \Delta\varphi_2 + \Delta\varphi_3$ is the Ramsey control phase, which is the sum of all relative phases. Here, $\Delta \varphi_2=\varphi_{L_2}-\varphi_{R_2}$ is the relative phase of the two $\pi/2$ pulses acting at sites $L_2$ and $R_2$.
Hence, we obtain from Eqs.~(\ref{eq:finalodd}) and (\ref{eq:finaleven}) the expectation value of the spin parity operator, which gives rise to an interference fringe as a function of the Ramsey control phase $\varphi$,
\begin{equation}
	\label{eq:spinparity}
	\langle \Pi \rangle {=} -\cos^2\hspace{-2pt}\left(\hspace{-2pt}\frac{\varphi_\text{ex}{-}\varphi}{2}\hspace{-2pt}\right)+\sin^2\hspace{-2pt}\left(\hspace{-2pt}\frac{\varphi_\text{ex}{-}\varphi}{2}\hspace{-2pt}\right) {=} {-} \cos(\varphi_\text{ex}{-}\varphi).
\end{equation}
By precisely varying the Ramsey control phase $\varphi$ (e.g., scanning $\Delta\varphi_3$, see Sec.~\ref{sec:posdeppulses}), one records a Ramsey fringe whose position directly reveals the exchange phase $\varphi_\text{ex}$.
Note that, as already mentioned, dynamical phases cancel out (see Sec.~\ref{sec:decoherence}) and, thus, have no effect on the fringe position.

For sake of completeness, it is worth discussing also the evolution of the state in Eq.~(\ref{eq:transportevolutionsamesite}), for which the two atoms occupy the same site.
At both sites $L_2$ and $R_2$, the $\pi/2$ pulse realizes the equivalent of a beam splitter, which mixes the paths of two indistinguishable photons in an Hong-Ou-Mandel interferometer. After the pulse it results that
bosonic atoms ($\varphi_\text{ex}=0$) bunch in the same spin state, whereas fermionic ones ($\varphi_\text{ex}=\pi$) end up in different spin states.
In both cases, the final result does not depend on the Ramsey control phase $\varphi$ because local measurements (i.e., measurements acting either on atoms at sites $L_2$ or $R_2$) are insensitive to $\varphi$.

\subsection{Position-dependent pulses}
\label{sec:posdeppulses}
Position-dependent pulses can be realized with optical addressing of the individual sites, either using tightly focused Raman laser beams 
or microwave radiation in conjunction with auxiliary Stark-shifting laser beams \cite{Weitenberg11}.
High fidelities above $\SI{99}{\percent}$ and small cross-talk errors below $\SI{1}{\percent}$ should be reachable \cite{Xia:2015} since the distance between the addressed sites, which is comparable to the initial separation between the two atoms, can be chosen significantly larger than the wavelength of the addressing light field.
As an alternative to optical addressing, one can use Zeeman addressing through a magnetic field gradient \cite{Karski:2010}, which in principle can even allow sub-wavelength resolutions.

As shown in Sec.~\ref{sec:RamseyFringe}, position-dependent pulses allow one to control the Ramsey fringe position by varying any of the relative phases $\Delta\varphi_1$, $\Delta\varphi_2$, $\Delta\varphi_3$.
However, position-dependent pulses are strictly necessary only for the $\pi$ pulses applied at sites $L_3$ and $R_3$, which allow the paths to be recombined as shown in Fig.~2 of the main text.

\begin{figure}[t]
\includegraphics[width=1\columnwidth]{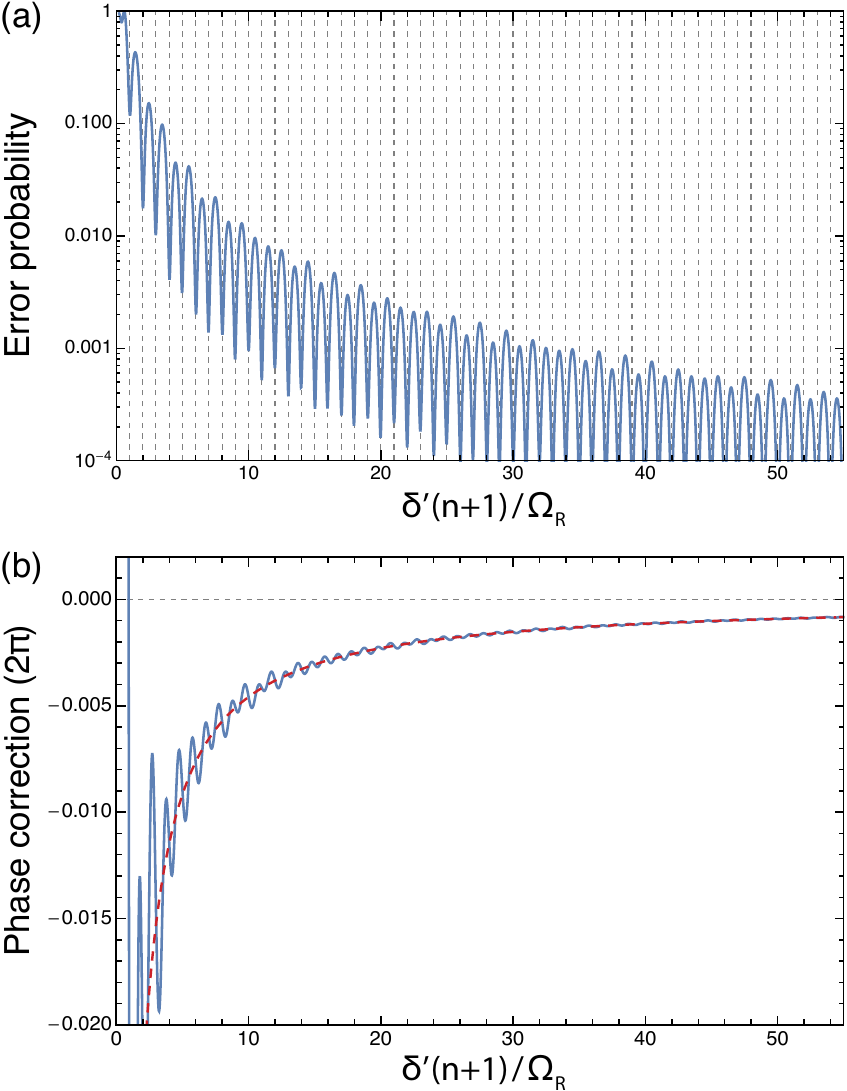}
\caption{\label{fig:position-resolved-pulses}
Analysis of imperfections resulting from the $\pi$ pulse with Zeeman addressing of $L_3$ and $R_3$ sites.
The results are obtained by solving numerically the pseudo-spin evolution with $n=10$.
The quantities displayed as a function $\delta'(n+1)/\Omega_\text{R}$ show a negligible dependence on the initial distance between atoms $n$ when $n\gtrsim 10$.
(a) Probability $p_\text{err}$ that spurious spin components are produced by the $\pi$ pulse, resulting in atoms not reaching the detection sites $L_2$ and $R_2$.
(b) Phase correction shifting the Ramsey fringe obtained by exact numerical computation (blue solid line) and by ac-microwave shift $\Delta\varphi_{\text{MW}}$ (red dashed line).
The Zeeman contribution $\Delta\varphi_\text{Z}$ to the phase shift has been subtracted from the displayed curves for better visibility.
\vspace{0.7cm}
}
\end{figure}

Importantly, errors in the spin populations produced by the first $\pi/2$ pulse and by the intermediate $\pi$ pulse do not affect the visibility of the two-atom Ramsey interferometer because the spurious spin components are not transported to the detection sites $L_2$ and $R_2$ and, thus, do not contribute to the final parity signal.
Their only effect is a reduction of the signal-to-noise ratio for a fixed number of atom pairs probed.
However, errors in the spin populations produced by the last $\pi/2$ pulse cannot instead be discarded by post-selection since all spin components at sites $L_2$ and $R_2$ contribute to the parity signal.

In what follows, we discuss three examples of pulse errors affecting the spin populations:
(\textit{i}) We assume that the first Ramsey pulse differs from a $\pi/2$ pulse, and rotates (in the Bloch sphere representation) the pseudo-spin by an angle $\theta$ at both sites $L_1$ and $R_1$.
It follows that a fraction equal to $1-\cos^2(\theta/2)\sin^2(\theta/2)=1-\sin^2(\theta)/2$ of atom pairs does not contribute to the parity signal because, in these events, the two atoms do not reach the final detection sites.
The minimum fraction of discarded pairs by post-selection amounts to $1/2$ and is obtained when $\theta=\pi/2$.
(\textit{ii}) We here assume that the impaired pulse is the intermediate pulse, which rotates the pseudo spin by an angle $\theta_\text{out}$ at the outermost sites, and by an angle $\theta_\text{in}$ at the innermost sites.
In this case, the fraction of pairs not contributing to the parity signal amounts to $1-\sin^2(\theta_\text{out}/2)\cos^2(\theta_\text{in}/2)/2$, which reaches the minimum value of $1/2$ for $\theta_\text{out}=\pi$ and $\theta_\text{in}=0$.
(\textit{iii}) We assume that the last Ramsey pulse is impaired, rotating the pseudo spin by an angle $\theta$ at both sites $L_2$ and $R_2$.
In this case, the parity signal obtained from post-selected pairs detected at $L_2$ and $R_2$ is $\langle\Pi\rangle=-\cos^2(\theta)-\sin^2(\theta)\cos(\phi) =\Pi_0 -\mathcal{V}\cos(\phi)$.
This interference fringe has a reduced visibility $\mathcal{V}=\sin^2(\theta)$ and a vertical offset of  $\Pi_0=-\cos^2(\theta)$.
For $\theta=\pi/2$, the visibility is maximum $\mathcal{V}=1$ and the offset vanishes.

In the remainder of this section, we consider in greater detail errors produced by the intermediate $\pi$ pulse when Zeeman addressing is employed \cite{Karski:2010}.
While optimal pulse shaping can reduce errors, for simplicity we assume here two square pulses with slightly different frequencies so that, in the presence of a magnetic field gradient, each of them resonantly addresses either $L_3$ or $R_3$.
Without loss of generality (see Sec.~\ref{sec:decoherence}), we assume that the linear detuning, $\delta' x$, induced by the magnetic field gradient, is symmetric with respect to the central position $x=0$ between the two atoms. Here, $\delta'$ is the angular frequency shift per lattice site, and $x$ is the position in units of lattice sites.
In this notation, $L_3$ and $R_3$ have positions $x=\pm(n+1)$, whereas $L_1$ and $R_1$ have positions $x=\pm1$.
In experiments, the detuning per lattice site $\delta'$ can be of the order of $(2\pi)\SI{15}{\kilo\hertz}$ \cite{Karski:2010}.

We have numerically computed the pseudo-spin evolution for the Zeeman-addressed $\pi$ pulse.
The error probability $p_\text{err}$ to produce spurious spin components is shown in Fig.~\figref{fig:position-resolved-pulses}{a} as a function of $\delta'(n+1)/\Omega_{\text{R}}$.
As explained above, these errors do not affect the visibility of the parity signal, but reduce the signal-to-noise ratio for a fixed number of pairs probed in the experiment.
The overall fraction of discarded pairs by post-selection amounts to $1/2+p_\text{err}/2$.
The numerical results show a strong reduction of the error probability when the ratio $\delta'(n+1)/\Omega_R$ is an integer number.
The error suppression occurs when the two $\pi$ pulses, which are resonant with the outermost sites, produce a vanishing rotation at the innermost sites, $L_1$ and $R_1$; a similar error suppression is demonstrated in Ref.~\cite{Xia:2015}, however, in a slightly different configuration where only a single $\pi$ pulse is employed.

If we choose the error minimum at $\delta'(n+1)/\Omega_R=2$, we expect from Fig.~\figref{fig:position-resolved-pulses}{a} a negligibly small error probability $p_\text{err}<\SI{2}{\percent}$. Assuming a microwave-induced bare Rabi frequency of $\Omega_\text{R}=(2\pi) \SI{60}{\kilo\hertz}$ \cite{Karski:2010}, we obtain, for an initial spatial separation of $n=10$ sites, $\delta'\approx (2\pi) \SI{10}{\kilo\hertz}$, which is within the reach of experiments.
Smaller Rabi frequencies allows one to use weaker magnetic field gradients, but also imply longer spin-flip times and, thus, an increased rate of decoherence-induced pulse errors.

For the interferometric measurement of the exchange phase $\varphi_\text{ex}$, it is important that phase errors are minimized.
As discussed in Sec.~\ref{sec:RamseyFringe}, the fringe position depends on the relative phase $\Delta\varphi_3$ between the two pulses $\pi$ acting at sites $L_3$ and $R_3$.
However, the magnetic field gradient adds a Zeeman contribution to the phase shift, $\Delta \varphi_\text{Z} = -[(n+1)-1] \pi \delta'/\Omega_R = -n \pi \delta'/\Omega_R$.
This contribution can be compensated, see Sec.~\ref{sec:decoherence}, by symmetrizing in time the interferometric scheme.

To discern phase corrections beyond the two leading contributions from $\Delta \varphi_3$ and $\Delta\varphi_\text{Z}$, we compute the phase shift of the Ramsey fringe using the evolution of the pseudo spin computed numerically for the Zeeman-addressed $\pi$ pulse.
The computed phase shift is shown in Fig.~\figref{fig:position-resolved-pulses}{b} as a function of $\delta'(n+1)/\Omega_\text{R}$, after subtracting the two leading contributions.
The remaining phase shift exhibits an overall hyperbolic profile, with small oscillations on top of it.
The hyperbolic profile can be well explained as the result of the ac-microwave shift induced by the $\pi$ pulses on atoms situated at the innermost sites $L_1$ and $R_1$.
This shift can be computed with second-order perturbation theory, yielding
\begin{multline}
	\label{eq:microwaveshift}
	\Delta \varphi_\text{MW}= \frac{\Omega_\text{R}^2}{4}\bigg(
	\frac{1}{\delta'-\omega_\mathrm{R_3}} {+} \frac{1}{\delta'-\omega_\mathrm{L_3}} {-} \frac{1}{-\delta'-\omega_\mathrm{R_3}} \\{-} \frac{1}{-\delta'-\omega_\mathrm{L_3}}
	  \bigg)
	  \frac{\pi}{\Omega_\text{R}}=  -\frac{\pi(n+1)}{n(n+2)}\left[\frac{\delta'(n+1)}{\Omega_\text{R}}\right]^{-1}
\end{multline}
where  $\omega_\mathrm{{L_3},{R_3}} = \pm \delta'(n+1)$ are the resonance frequencies at sites $L_3$ and $R_3$.
Moreover, a closer inspection of Fig.~\figref{fig:position-resolved-pulses}{b} reveals that the phase correction computed numerically coincides with the ac-microwave shift  estimation in Eq.~(\ref{eq:microwaveshift}) when the error probability reaches a local minimum $p_\text{err}$ at integer values of $\delta'(n+1)/\Omega_\text{R}$.
Hence, we obtain that at these special values, the Ramsey fringe is shifted by $\Delta \varphi_3+\Delta\varphi_\text{Z} +\Delta\varphi_\text{MW}$.
In the two-atom Ramsey interferometer, $\Delta\varphi_3$ represents the precisely adjustable Ramsey control phase, $\Delta\varphi_\text{Z}$ cancels out due to spin-echo refocusing (see Sec.~SA.3), and the residual systematic shifts of the Ramsey fringe position of the order of \SI{1}{\percent} of $2\pi$ [see Fig.~\figref{fig:position-resolved-pulses}{b}] can be removed by subtracting the phase correction $\Delta \varphi_\text{MW}$ term.

\subsection{Dephasing analysis}
\label{sec:decoherence}
In single-particle interferometry, slow drifts of magnetic fields, magnetic field gradients, and imperfect control of dynamical phases during transport are often responsible for the loss of first-order coherence.
In the two-particle Ramsey interferometer sketched in Fig.~2 of the main text, these dephasing mechanisms could, in principle, also cause a reduction of the visibility of the parity signal and systematic errors in the measurement of the exchange phase $\varphi_\text{ex}$.
Here, we show that these mechanisms do not affect second-order coherence probed by the two-atom Ramsey interferometer and, hence, cause no shift of the measured Ramsey fringe.

To protect the interferometer against dephasing mechanisms, we require a few assumptions on the atom transport:
(1) Atoms follow spatially mirrored trajectories for the two states $\ket{{\uparrow}}$ and $\ket{{\downarrow}}$.
(2) For each given pseudo-spin state, all single-site transport operations are realized through the same shift of the periodic lattice potential \cite{Robens:2017}.

In the remainder of this section, we consider separately different mechanisms that could potentially cause dephasing. We compute the Ramsey phase shift as the difference between the phases accumulated by the innermost and outermost two-atom paths.

\paragraph{Fluctuating spin-independent forces}
We assume that an external force (e.g., gravity force) is directed along the lattice direction, acting on both spin components equally.
The force gives rises to a spin-independent linear potential gradient of the form $\hbar \omega_\text{B} (x-x_0)$, where $\omega_\text{B}$ is the Bloch frequency proportional to the force, $x$ is the position along the lattice in units of lattice sites, and $x_0$ is the position in the same units at which the potential vanishes.
The value of $x_0$ determines an overall energy shift equal to $-\hbar \omega_\text{B} x_0$, which cannot not influence the interferometric signal.
Thus, we can simply choose $x_0$ equal to the midpoint between the two atoms.

The overall phase acquired by the innermost two-atom path is zero because, based on assumption (1), the two atoms are at opposite positions in space at all times.
Likewise, the overall phase acquired by the outermost two-atom path is also zero.
Hence, the position of the Ramsey fringe is insensitive to fluctuations of a spin-independent force.
Because full phase cancellation occurs at all times, the scheme is also robust against fast fluctuations of the force.

\paragraph{Fluctuating magnetic fields} We assume that a homogeneous magnetic field is present (e.g., to define the quantization axis), causing a differential Zeeman energy shift for the two states $\ket{{\uparrow}}$ and $\ket{{\downarrow}}$.
We ignore any common-mode Zeeman energy shift because, if present, it only causes a spin-independent energy shift, which cannot affect the relative phase measured by the interferometer.

The overall phase acquired by the innermost two-atom path vanishes because the two atoms are in opposite spin states at all times.
For the same reason, the outermost two-atom path also yields a vanishing overall phase.
Hence, the position of the Ramsey fringe is insensitive to fluctuations of the magnetic field.
Because full phase cancellation occurs at all times, the scheme is also robust against fast fluctuations of the magnetic field.

\paragraph{Slowly fluctuating magnetic field gradients}
We assume that a magnetic field gradient is present along the lattice direction (as occurs for Zeeman-addressed $\pi$ pulses, see Sec.~\ref{sec:posdeppulses}), which is static on the timescale of the interferometric sequence, but can slowly fluctuate at longer times.
Due to the magnetic field gradient, the atoms experience a spin-dependent potential gradient of the form $\hbar \omega_{\text{B},\uparrow} (x-x_{0,\uparrow})\ket{{\uparrow}}\bra{{\uparrow}} - \hbar \omega_{\text{B},\downarrow} (x-x_{0,\downarrow})\ket{{\downarrow}}\bra{{\downarrow}}$, where $x$ is the position along the lattice, and  $\omega_{B,s}$,  $x_{0,s}$ are the Bloch frequency and the zero-crossing position for the two spin components $\ket{s}$, respectively.

Because fluctuations of $x_{0,s}$ are equivalent to fluctuations of a magnetic field, these cannot influence the Ramsey signal, as we have shown above.
Thus, we can conveniently choose $x_{0,\uparrow} = x_{0,\downarrow}$ equal to the midpoint between the two atoms.

The potential can be decomposed in the sum of a spin-dependent potential gradient, whose strength is proportional to  $(\omega_{B,\uparrow}+\omega_{B,\downarrow})/2$, and a spin-independent potential gradient, whose strength is proportional to  $(\omega_{B,\uparrow}-\omega_{B,\downarrow})/2$.
Because, as shown above, fluctuations of a spin-independent potential gradient do not affect the Ramsey signal, we can focus on the spin-dependent contribution by simply assuming $\omega_{B,\uparrow}=\omega_{B,\downarrow}$.

Because of assumption (2), the innermost two-atom path acquires a zero phase during the first $n$ shift operations, and a nonzero phase only during the last shift operation.
It turns out, however, that this phase equals that acquired by the outermost two-atom path.
In fact, relying again on assumption (2), the phase acquired by the outermost two-atom path during the last $n$ shift operations vanishes due spin-echo refocusing, and a nonzero phase is acquired only during the first shift operation.
Due to assumption (2), the nonzero phase acquired by the innermost path during the last shift operation equals that acquired by the outermost path during the last shift operation.
Hence, because their difference vanishes, we conclude that the position of the Ramsey fringe is insensitive to slow fluctuations of the magnetic field gradient.

\paragraph{Dynamical phases by atom transport}
Each single-site shift operation causes the atom to acquire a dynamical phase.
This dynamical phase includes both a kinetic and potential contribution, which for our purpose are not required to be separated.
In the most general case, we assume that the dynamical phase acquired by $\ket{{\uparrow}}$ and $\ket{{\downarrow}}$ states are different.

After a single-site shift operation, the dynamical phase acquired by the innermost two-atom path comprises the dynamical phase from state $\ket{{\uparrow}}$ and that from state $\ket{{\downarrow}}$.
Likewise, the dynamical phase acquired by the outermost two-atom path comprises the same dynamical phase from state $\ket{{\uparrow}}$ and that from state $\ket{{\downarrow}}$.
Hence, because the phase difference vanishes, we conclude that the Ramsey signal is insensitive to dynamical phases arising from atom transport.

\subsection{Partial indistinguishability}
Considering the scheme sketched in Fig.~2 of the main text, if the two atoms populate initially different vibrational states, one can discriminate whether the atoms have travelled the outermost or innermost paths by simply measuring the vibrational level of each atom at the end of the Ramsey interferometer.
Thus, the fact that the which-way information can be detected suppresses interference between the two particles:
A spin-parity measurement would then yield a constant signal equal to $\num{0}$, indicating the absence of correlations.

It is thus necessary to initially cool atoms to the lowest vibrational state in order to make them indistinguishable in the motional degree of freedom.
Further, it is also important that the atoms are transported without creating any vibrational excitation to avoid leaving a trace of the which-way information in the vibrational state; polarization-synthesized optical lattices enable spin-dependent transport with a virtually zero probability \mbox{$<\SI{1}{\percent}$} to create a vibrational excitation \cite{Robens:2016c}.
Hence, in the rest of this section we focus on the effect of imperfect cooling, which is at the present the dominant cause of a reduced visibility of the parity signal.

By starting the experiment from a Mott insulator state and retaining a small number of individual atoms (exactly two atoms for our purpose) at separate sites, a ground-state probability $p_0$ at around $\SI{95}{\percent}$ has been demonstrated \cite{Islam:2015}.
Sideband cooling techniques have so far only reached $p_0\lesssim \SI{90}{\percent}$ \cite{Kaufman:2012}, mainly due to small trap frequencies along at least one of the confining directions.
While we expect that ground-state probabilities close to unity will be achieved in the near future with sideband cooling techniques, it is interesting to consider the effect of imperfect cooling (i.e., $p_0<1$) on the visibility of the correlation signal $\langle\Pi\rangle$.

Thus, the spin parity signal $\langle \Pi \rangle$ depends on the probability, $\mathcal{P}_\text{indist}$, that the two atoms occupy the same vibrational state.
To compute it, we assume that the cooling procedure prepares a statistical mixture of vibrational states, denoted here $\rho^{(L)}$ and $\rho^{(R)}$ for the the atom on the left- and right-hand side, respectively.
With this assumption, one directly obtains that the probability for the two atoms to be indistinguishable in the motional degree of freedom is $\mathcal{P}_\text{indist} = \tr{(\rho^{(L)}\rho^{(R)})}$.
For two identical statistical mixtures $\rho^{(L)}=\rho^{(R)}$ (as is likely the case), note that $\mathcal{P}_\text{indist}$ corresponds to the purity of the initial single-atom states, which is 1 for perfect ground-state cooling (i.e., when $p_0=1$).
Hence, we conclude that the spin-correlation measurement yields
\begin{eqnarray}
\label{eq:signal_finite_temp}
\langle \Pi \rangle&=& (1-\mathcal{P}_\text{indist})\cdot 0 - \mathcal{P}_\text{indist}\cdot\cos(\varphi-\varphi_\text{ex})\nonumber\\
&=&-\mathcal{V}\cos\left(\varphi-\varphi_\text{ex}\right)\,,
\end{eqnarray}
where $\mathcal{V}=\mathcal{P}_\text{indist}$ defines the visibility of the parity signal. The expression in Eq.~(\ref{eq:signal_finite_temp}) makes use of the spin-parity signal for indistinguishable atoms, which is derived in Eq.~(\ref{eq:spinparity}).

To get insight into Eq.~(\ref{eq:signal_finite_temp}), we assume for both statistical mixtures a thermal distribution in a harmonic trap, with trapping frequencies $\omega_x$, $\omega_y$, and $\omega_z$. Denoting by $p_{0,i}$ the probability of occupying the lowest vibrational state along the $i$-axis (these can be three different values in experiments), it follows that $p_0 = p_{0,x}p_{0,y}p_{0,z}$ and that
\begin{equation}
\mathcal{V} = \frac{p_0}{(2-p_{0,x})(2-p_{0,y})(2-p_{0,z})}\,.
\end{equation}
If we assume isotropic cooling with $p_0=p_{0,i}^3$, and that $p_0\approx 1$, we obtain that $\mathcal{V}\approx p_0^2$ up to corrections of the order of $\mathcal{O}[(1-p_0)^2]$. This simple formula shows that the visibility is in good approximation equal to the probability that both atoms populate the lowest vibrational state.
For example, we obtain $\SI{80}{\percent}$ visibility for $p_0\approx \SI{90}{
\percent}$, and $\SI{50}{\percent}$ visibility for $p_0 \approx \SI{70}{
\percent}$.

\begin{figure}[t]
\includegraphics[width=1\columnwidth]{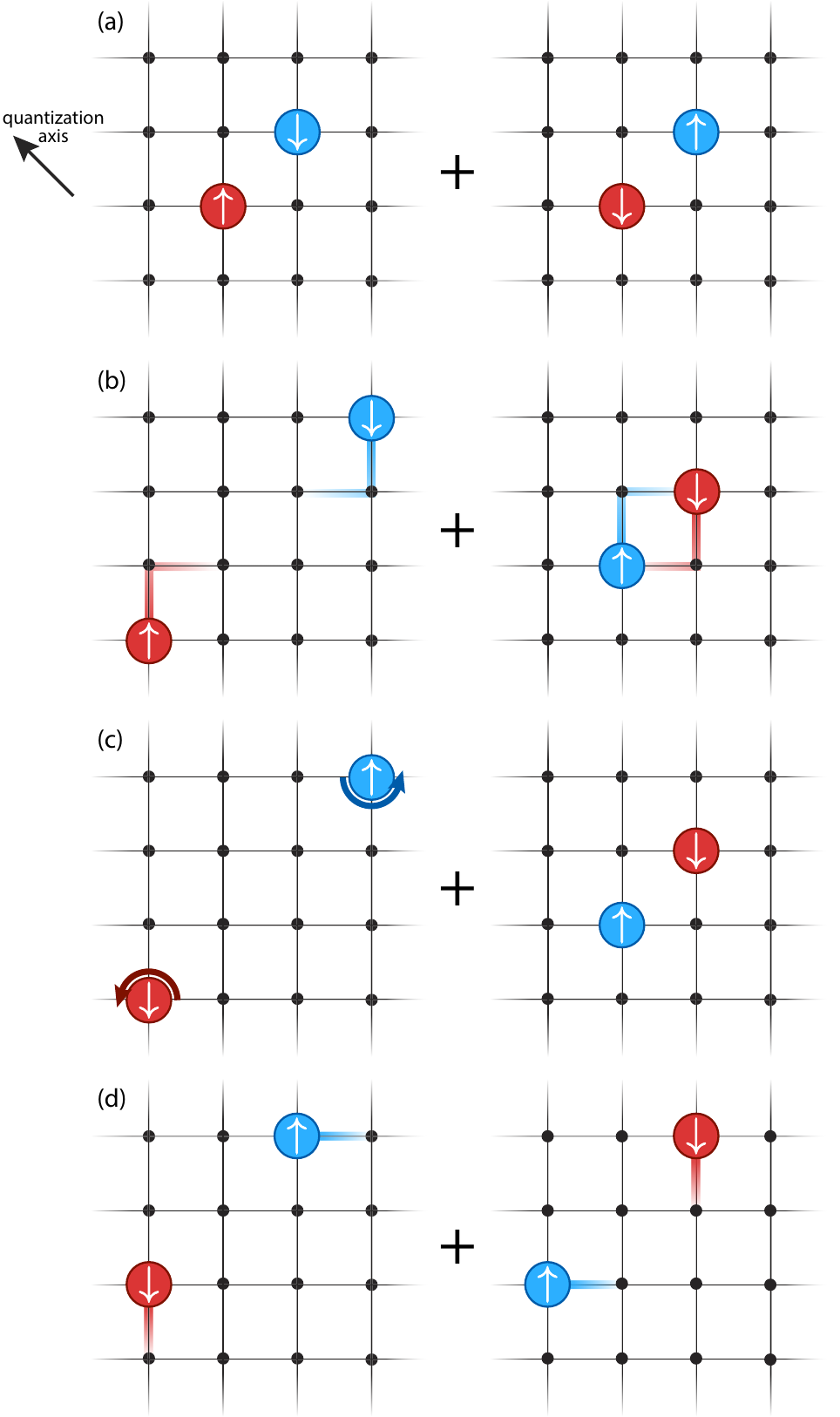}
\caption{\label{fig:2Dscheme}Two-atom Ramsey sequence on a two-dimensional spin-dependent lattice. Only the evolution of the nonseparable $\ket{{\Psi_\text{odd}}}$ state, see Eq.~(\ref{eq:psioddinit}), is displayed ($+$ sign indicates the superposition of the two states). The different panels represent the state: (a) after an initial $\pi/2$ pulse, (b) after a L-shaped spin-dependent shift, (c) after a spin flip operation affecting the outermost region only, (d) after a straight
spin-dependent displacement. The sequence is completed by applying a final $\pi/2$ pulse and by measuring the spin state of both atoms (not shown). The quantization axis along the diagonal \cite{Groh:2016} is also shown.}
\end{figure}

\subsection{Two-dimensional scheme}
\label{sec:twodim}
Here, we briefly discuss a protocol for neutral atoms confined in a two-dimensional spin-dependent optical lattice. This can be realized with polarization-synthesized (PS) optical lattices, which have recently been demonstrated in one dimension \cite{Robens:2017}. The scheme extending PS optical lattices to two-dimensions is described in Ref.~\citenum{Groh:2016}. A sketch of the two-dimensional protocol is shown in Fig.~\ref{fig:2Dscheme}. The main difference between this protocol and that presented in Fig.~2 of the main text is that here the two atoms are exchanged, see panel (b), without never crossing each other. Hence, the two-dimensional protocol ensures that the two atoms never interact with each other.
Through an analysis of dephasing mechanisms similar to that presented in Sec.~\ref{sec:decoherence}, one can show that the scheme sketched in the figure is insensitive to fluctuations of spin-independent forces, fluctuations of magnetic fields, fluctuations of magnetic field gradients, and imperfect control of dynamical phases during transport.
In the case the quantization axis, defined by a static magnetic field of a few gauss, is chosen along the diagonal direction shown in  Fig.~\ref{fig:2Dscheme}, the two-dimensional two-atom Ramsey interferometer is even robust against fast fluctuations of magnetic field gradients, in contrast to the one-dimensional scheme, which is only robust against slow fluctuations, see Sec.~\ref{sec:decoherence}.
Note that if the quantization axis is oriented along the other diagonal, the transport scheme projected along the quantization axis is equivalent to the one-dimensional scheme sketched in Fig.~2 of the main text.
In this configuration, the two-dimensional scheme is only robust against slow fluctuations of magnetic field gradients like the one-dimensional scheme.

\subsection{Choice of atomic species}
The implementation scheme illustrated in Fig.~2 of the main text makes use of state-dependent optical lattices to swap two atoms and measure their exchange phase in an interferometric scheme.
While other schemes relying on different realizations of state-dependent potentials (e.g., rf-dressed or microwave-dressed potentials \cite{Boehi:2009}) are in principle conceivable, our proposed scheme based on state-dependent potentials has the important property of being robust against dephasing mechanisms, as explained in Sec.~\ref{sec:decoherence}, and, moreover, can be realized with currently available technology \cite{Mandel:2003,Steffen:2012,Robens:2017}.
However, the usage of state-dependent optical potentials also has a downside, as it limits the number of suitable atomic species for such an experiment.

For rubidium and cesium atoms, which have been employed in previous state-dependent-transport experiments  \cite{Mandel:2003,Steffen:2012,Robens:2017}, all stable isotopes are bosonic.
Lighter alkali atoms like potassium or lithium, which possess both fermionic and bosonic isotopes, are not suitable for state-dependent transport because the spin-orbit coupling intrinsic to the atom is too weak, resulting in a small fine-structure splitting; in fact, the small splitting prevents state-dependent optical potentials without causing too strong off-resonance scattering of photons.

However, state-dependent transport could be realized with fermionic atoms as well, by exploiting the nuclear spin (hyperfine interaction) instead of the electron spin (spin-orbit coupling), according to the scheme for alkaline-earth-metal atoms (e.g., Yb or Sr atoms) proposed in Ref.~\cite{Daley:2011}.
Only fermionic isotopes are suited for this scheme, though, because of their nonzero nuclear spin, while all stable bosonic isotopes have zero nuclear spin.

To our best knowledge, aluminum is the only atomic species that allows the exchange of both fermionic ($^{26\!}$Al) and bosonic ($^{27\!}$Al) isotopes with the same experimental setup.
One of the authors (A.A.) considers realizing state-dependent potentials for aluminum atoms by exploiting the inherent spin-orbit coupling of the ground-state p orbital.
In this case (more generally, for group III atoms), the nonvanishing spin-orbit coupling in the electronic ground state makes it possible to choose the wavelength of the optical-lattice laser far detuned from any excited electronic state.

Such a setup would allow one to compare the absolute (i.e., calibration-free) value of the exchange phase measured for bosons and fermions with the same setup.
In fact, the measurement scheme for the detection of the exchange phase is insensitive to isotopic differences in the mass or in the ac-Stark shifts:
Different masses, as well as different ac-Stark shifts, would lead to different dynamical phases during state-dependent transport operations for the single-atom quantum paths, and yet would not affect the relative phase between the two two-atom quantum paths interfering in the two-atom Ramsey interferometer.
As explained in Sec.~\ref{sec:decoherence}, these two-atom paths acquire exactly the same dynamical phase.

\section{Protocol for trapped ions}

In this section, we describe a protocol to extract the wave-function symmetry of two trapped ions. We first give arguments based on the symmetry of a suitable Hamiltonian exchanging the two ions, and we describe how this Hamiltonian can be realized exploiting the micromotion in a linear Paul trap. Then, we present numerical results suggesting that the ion exchange can be performed adiabatically, and we discuss possible experimental imperfections.

\subsection{Symmetry properties of the two-ion wave function}
Here, we extend simple symmetry arguments how an experiment can make the exchange symmetry of a pair of trapped ions visible, even if the wave functions of the ions never overlap.
We assume a pair of identical charged particles, trapped in the radial ($x$-/$y$-) plane of a linear Paul trap by an axial confinement (in the $z$-direction) that is stronger than the radial one.  To keep the notation simple, we will drop the part describing the axial confinement from all formulas given in the following. The positions of the ions are at $z_\ell=0$ and  $\bm{r}_\ell=\left(x_\ell,y_\ell\right)$, $\ell=1,2$. In the secular approximation, the trap provides a harmonic potential given by
\begin{equation}
U_{\text{trap}}=\frac{1}{2}m(\omega_x^2(x_1^2+x_2^2)+\omega_y^2(y_1^2+y_2^2)).
\end{equation}
In addition, the ions interact via the Coulomb interaction 
\begin{equation}
V_{\text{Coul}}=\frac{e^2}{4\pi\epsilon_0|\bm{r}_1-\bm{r}_2|}\,.
\end{equation}
For degenerate radial trapping frequencies,  $\omega_x=\omega_y$, the potential is rotationally symmetric.
However, the ions experience micromotion in the radial plane, which breaks rotational symmetry. As Fig.~3 of the main text shows, depending on the orientation of the crystal the direction of micromotion may be aligned with the crystal axis or perpendicular to it. Similar to the secular approximation used to calculate the oscillation frequencies of a trapped ion in a time-dependent quadrupolar field, the effective dynamics of the ions is described by averaging their Coulomb energy over one period of the rf-driving field. As shown further below, this process leads to a modification of the Coulomb potential
\begin{eqnarray}
\overline{V}_{\text{Coul}}&=&\langle\frac{e^2}{4\pi\epsilon_0|\bm{r}_1(t)-\bm{r}_2(t)|}\rangle_T\nonumber\\
 &=& \frac{e^2}{4\pi\epsilon_0|\bm{r}_1-\bm{r}_2|}\left(1+\frac{3}{16}q^2\cos^2(2\theta)\right)
\end{eqnarray}
that breaks the rotational symmetry even in the case where $\omega_x=\omega_y$. Here, $\langle\bullet\rangle_T$ denotes time averaging, $q$ denotes the ion trap's $q$-factor, and $\theta$ the direction in the radial plane ($\theta=0$ and $\theta=\pi/2$ being the $x$- and $y$-direction).  

We assume that initially $\omega_y>\omega_x$ so that the ion crystal is aligned along the x-direction with ion 1 sitting at position $\bm{r}_0$ and ion 2 at position $-\bm{r}_0$. We further assume that the initial two-ion wave function is a product state
\begin{equation}
\Psi_i=\psi_{\bm{r}_0}(\bm{r}_1)\psi_{-\bm{r}_0}(\bm{r}_2)\label{eq:psii}
\end{equation}
where $\psi_A(\bm{r})$ describes a Gaussian wave packet centered around position $\bm{r}_A$.  
For a pair of identical particles, the constraints imposed by quantum statistics need to be considered. In the following, we will assume that both particles are prepared in the same spin state. As a consequence, the spatial part of the wave function will be symmetric for bosons and antisymmetric for fermions. If the initial state (\ref{eq:psii}) is (anti-)symmetrized, it becomes
\[
\tilde{\Psi}_i=(\psi_{\bm{r}_0}(\bm{r}_1)\psi_{-{\bm{r}_0}}(\bm{r}_2))+(-1)^\sigma(\psi_{-{\bm{r}_0}}(\bm{r}_1)\psi_{{\bm{r}_0}}(\bm{r}_2))
\]
with $\sigma=0$ for bosons and $\sigma=1$ for fermions. Here and later on, a tilde sign will denote (anti-)symmetrized wave functions. 

Due to the symmetries of the problem, it is more convenient to express the ion positions in terms of their center-of-mass vector $\bm{R}=(X,Y)$ and relative position vector $\bm{r}=(x,y)$,
\begin{eqnarray}
\bm{r}_1&=&\bm{R}+\bm{r}/2\label{eq:coordinatesC}\\
\bm{r}_2&=&\bm{R}-\bm{r}/2\label{eq:coordinatesR}.
\end{eqnarray}
In terms of these variables, the Hamiltonian separates into $H=H_c+H_r$ where 
\begin{eqnarray}
H_c&=&-\frac{\hbar^2}{2M}\nabla_c^2+\frac{1}{2}M(\omega_x^2X^2+\omega_y^2Y^2)\label{eq:Hc}\\
H_r&=&-\frac{\hbar^2}{2\mu}\nabla_r^2+\frac{1}{2}\mu(\omega_x^2x^2+\omega_y^2y^2) + \overline{V}_{\text{Coul}}\label{eq:Hr}
\end{eqnarray}
with $M=2m$ and $\mu=m/2$ the reduced mass. 

Written in terms of the center-of-mass and relative coordinates, 
the initial state $\Psi_i$ also factors into a product of wave functions,
\begin{equation}
\Psi_i=\psi_c(\bm{R})\psi_\text{rel}(\bm{r}/2-\bm{r}_0).
\end{equation}
For trapped ions, this formulation correspond to a description in terms of the ion normal modes of motion. 

In these coordinates, the (anti-)symmetrization amounts to the operation
\begin{eqnarray*}
\bm{R}=\frac{1}{2}(\bm{r}_1+\bm{r}_2)&\rightarrow& \phantom{(-1)}\bm{R}\\
\bm{r}=\phantom{\frac{1}{2}}(\bm{r}_1-\bm{r}_2)&\rightarrow& (-1)^\sigma\bm{r}
\end{eqnarray*}
resulting in the (anti-)symmetrized wave function
\[
\tilde{\Psi}_i=\psi_c(\bm{R})\left(\psi_\text{rel}(\bm{r}/2-\bm{r}_0)+(-1)^\sigma\psi_\text{rel}(-\bm{r}/2-\bm{r}_0)\right).
\]
The dynamics of the protocol is governed by Hamiltonian $H$ given in eqs.~(\ref{eq:Hc}), (\ref{eq:Hr}) with changing trap frequencies. As the two parts of the Hamiltonian act respectively only on the center-of-mass and the relative coordinates, the wave function will remain in a product state. Due to this factorization, the bosonic or fermionic character of the particles only appears in the part involving the relative position, and we will drop the center-of-mass wave function $\psi_c$ in the following description.
\begin{figure*}[t]\centering
\includegraphics[width=\linewidth]{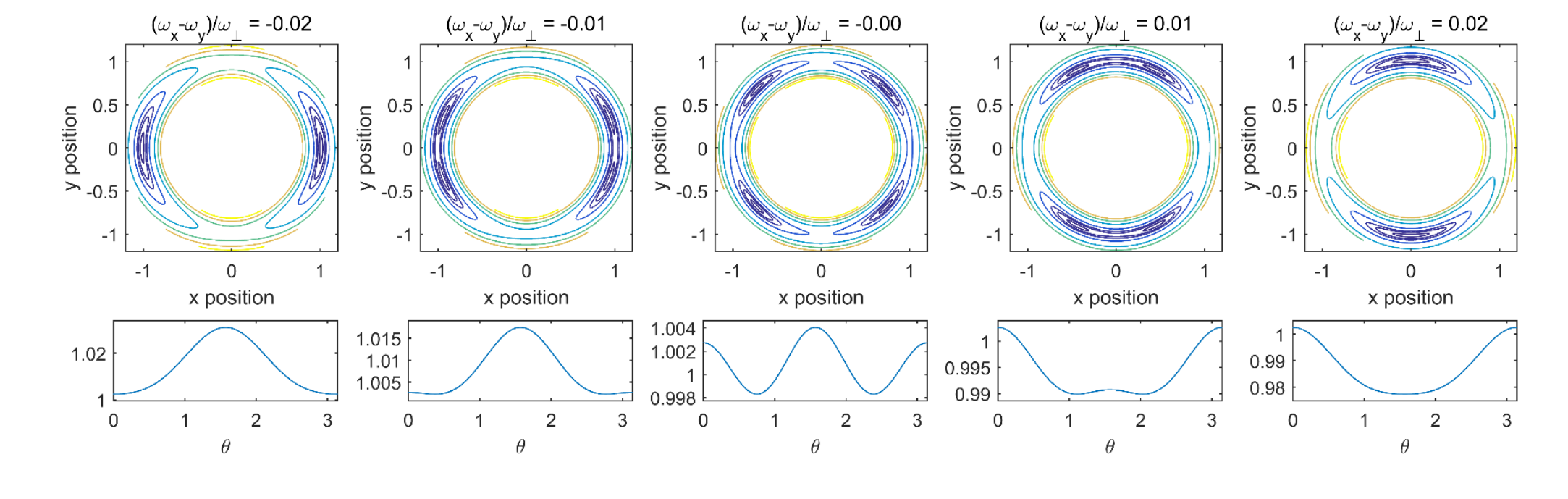}
\caption{\label{fig:timeaveragedPot} Numerical calculation of a time-averaged potential for the ions' relative motion (in units of the ions' equilibrium distance) for a $q$-parameter of $q=0.2$ and normal mode splittings. The leftmost and rightmost plot correspond to the anisotropy for which the rocking mode potential becomes quartic. The center column corresponds to an isotropic external potential. The plots in the lower line show the minimum potential as a function of the angle $\theta$.} 
\end{figure*}

In the proposed protocol, we assume that the anisotropy of the initial trapping potential is slowly changed from $\omega_x<\omega_y$ to $\omega_x>\omega_y$. 
The symmetry properties of the Hamiltonian (\ref{eq:Hr}) governing the relative motion, which are retained during this adiabatic change, permit us extend a simple argument to distinguish bosons and fermions: its total potential energy $V(x,y)=V_{\text{trap}}(x,y)+\overline{V}_{\text{Coul}}(x,y)$, 
fulfils
\[
V(x,y)=V(-x,y)=V(x,-y).
\] 
Therefore, if a wave function is initially symmetric or antisymmetric under $x\rightarrow -x$ or $y\rightarrow -y$, this symmetry will be conserved when the wave function is evolved under $H_r$. Now, assume a two-ion crystal of identical fermions orientated along the $x$-axis and prepared in the ground state of the rocking mode, which describes the out-of-phase ion motion normal to the crystal orientation, i.e., oscillates along $y$. This ion crystal has the fermionic antisymmetry plus the symmetry of the motional wave function, 
\begin{eqnarray*}
\psi_\text{rel}(x,y)&=&-\psi_\text{rel}(-x,-y)\;\;\;  \mbox{(fermionic antisymmetry)}\\
\psi_\text{rel}(x,y)&=&\phantom{-}\psi_\text{rel}(x,-y)\;\;\; \mbox{(symmetry of even Fock states)}
\end{eqnarray*}
and therefore also has the property 
\[
\psi_\text{rel}(x,y)=-\psi_\text{rel}(-x,y).
\]
As mentioned above, this symmetry will be conserved in the process in which the crystal is reoriented along the $y$-direction, and for this reason, the resulting wave function has no overlap with even Fock states of the rocking mode (which is now oriented along $x$). This argument shows that, because of the symmetries of the Hamiltonian, the initial rocking mode ground state cannot be mapped to the rocking mode ground state of the rotated crystal if we are dealing with fermions. When applying the same argument to a pair of bosons, one finds that the ground state is mapped to a wave function which does not have any overlap with the odd Fock states of the final crystal orientation.

The description of the rotation process can be further simplified, and made more explicit, by noting that in the process of lowering the radial anisotropy, the rocking mode frequency will become much lower than the stretch mode frequency. It seems then reasonable to assume that the motional wave function will approximately separate between both modes.  
As the stretch mode is always oriented along the direction of the crystal and the rocking mode normal to it, this translates into the product form
\[
\psi_\text{rel}(\bm{r}/2-\bm{r}_0)\approx f(|r|)\chi(\theta-\theta_0)
\]
if $\bm{r}=(r\cos\theta, r\sin\theta)$ and $\theta_0$ is the angle under which the crystal is aligned in the radial plane. For the variables $r$, $\theta$,  the (anti-)symmetrized wave function becomes
\[
\tilde{\psi}_\text{rel}(\bm{r}/2-\bm{r}_0)\approx f(|r|)\left(\chi(\theta-\theta_0)+(-1)^\sigma\chi(\theta+\pi-\theta_0)\right).
\]
As this relation shows, the symmetry properties of the wave function are carried by its angular part. During the adiabatic deformation of the time-averaged potential, this part will be modified. The upper line of Fig.~\ref{fig:timeaveragedPot} shows the potential energy as given in eq.~(\ref{eq:Hr}) for various anisotropies of the trapping potential, and its lower line shows the resulting angular potential. 
For $\omega_x-\omega_y<-\delta\omega_{crit}$, the ion crystal is aligned along the $x$-axis ($\theta_m=0$). The left column shows the case  ($\omega_x-\omega_y=-\delta\omega_{crit}$) where the time-averaged Coulomb potential makes the potential seen by the rocking mode quartic. For smaller anisotropies, the potential in the tangential direction breaks up into a double well. The well centers move from $\theta_m=0$ to positions $\theta_m = \pm\pi/4$ for $\omega_x=\omega_y$. For $\omega_x>\omega_y$ they move further outwards and finally, when the potential becomes quartic again ($\omega_x-\omega_y=\delta\omega_{crit}$), merge with a second set of minima at locations  $\theta_m = \pm\pi/2$.

How is the angular part of the wave function transformed in this process? We assume that the rocking mode is initially cooled to the ground state such that $\chi(\theta)\sim\exp(-\gamma\theta^2)$ is given by a Gaussian function. When the harmonic well splits into a double-well potential, $\chi(\theta)$  will be transformed into an even function that can be written as a sum of Gaussians, $\chi(\theta-\theta_m)+\chi(\theta+\theta_m)$, whose centers will move to $\theta_m = \pi/4$ for $\omega_x=\omega_y$. The initial (anti-)symmetrized wave function 
\[
\tilde{\psi}_i=  \chi(\theta)+(-1)^\sigma\chi(\theta+\pi)
\]
gets transformed in this first step into a wave function describing a superposition of crystal orientations,
\begin{eqnarray*}
\tilde{\psi}_{sup}&\sim& \left(\chi(\theta+\frac{\pi}{4})+\chi(\theta-\frac{\pi}{4})\right)\\
&&+(-1)^\sigma \left(\chi(\theta-\frac{3\pi}{4})+\chi(\theta+\frac{3\pi}{4})\right)\\
&=&\left(\chi(\theta+\frac{\pi}{4})+(-1)^\sigma \chi(\theta+\frac{3\pi}{4})\right)\\
&&+(-1)^\sigma \left(\chi(\theta-\frac{3\pi}{4})+(-1)^\sigma\chi(\theta-\frac{\pi}{4})\right),
\end{eqnarray*}
where in the last equation the terms are rearranged in such a way as to show that the superposition of the parts located at $\theta=\pi/4$ and $\theta=3\pi/4$ is symmetric for bosons and antisymmetric for fermions. Therefore, when these two parts are recombined into a single potential well at $\theta=\pi/2$, the final wave function will be
\[
\tilde{\psi}_f = \chi^{(\sigma)}(\theta-\frac{\pi}{2})+(-1)^\sigma\chi^{(\sigma)}(\theta+\frac{\pi}{2}),
\]
i.e., the uneven superposition of the fermionic case will be converted into the $n=1$ Fock state of the potential at $\theta=\pi/2$, denoted by $\chi^{(1)}(\theta)\sim\theta\exp(-\gamma\theta^2)$, whereas the even superposition of the bosonic case will be converted to the $n=0$ ground state of the potential [$\chi^{(0)}(\theta)=\chi(\theta)$].

\subsection{Time-averaged effective two-ion Coulomb interaction}
We now present the derivation of the effective potential $\overline{V}_{\text{Coul}}$ used above. 
In a two-ion crystal that is oriented in the radial plane of a linear rf-trap, as assumed above, the ions experience a fast micromotion. As a consequence, their potential energy obtains a periodic time dependence that can give rise to a modification of the normal mode frequencies of an ion crystal \cite{Landa:2012, Kaufmann:2012}. 

However, the influence of micromotion on the ions becomes much more drastic if the radial secular frequencies of the trap are degenerate or close to degeneracy. The reason is that even if the secular potential becomes radially isotropic for $\omega_x=\omega_y$, the micromotion electric field remains anisotropic as shown in Figure~3 of the main text: depending on the orientation of the ion crystal, the micromotion direction may be parallel to crystal axis or perpendicular to it. In the former case, time averaging the Coulomb interaction over the ion trajectory increases the average Coulomb energy (the increase in Coulomb energy at the inner turning point is higher than the decrease at the outer turning point). In the latter case, the micromotion slightly increases the average distance between the ions and therefore lowers the average Coulomb energy. Therefore, for a two-ion crystal, the time-averaged potential is no longer isotropic in the radial plane, an effect that was noticed already more than twenty years ago \cite{Moore:1994,Hoffnagle:1995}.

An ion will oscillate around the equilibrium position $(x_\ell,y_\ell)$ in the radial plane on a trajectory described by $x_\ell(t)=x_\ell (1+\frac{q}{2}\cos\Omega t)$ and  $y_\ell(t)=y_\ell(1-\frac{q}{2}\cos\Omega t)$ where $\Omega$ is the trap drive frequency and $q$ the $q$-parameter. Taylor-expanding the Coulomb energy up to second order in $q$ results in a time-averaged Coulomb potential equal to
\begin{eqnarray}
\overline{V}_{\text{Coul}}&=&\frac{e^2}{4\pi\epsilon_0}\langle\frac{1}{|\bm{r}_1(t)-\bm{r}_2(t)|}\rangle_T\nonumber\\
&=&\frac{e^2}{4\pi\epsilon_0|\bm{r}|}\left(1+\frac{q^2}{16}\left(3\cos^2(2\theta)-1\right)\right).\label{eq:AveragedCoulombPot}
\end{eqnarray} 
The time-averaged potential depends on the orientation $\theta$ of the ion crystal in the radial plane to second order in $q$. If the transverse oscillation frequencies of the trap are sufficiently different, the modification of the Coulomb potential will slightly weaken the rocking mode frequency in addition to very slightly increasing the equilibrium distance between the ions.
 However, if the trapping potential has nearly degenerate oscillation frequencies, the modified potential can make the curvature of the rocking mode's potential zero or even negative, thus destabilizing the orientation of the ion crystal.

\subsection{Modification of the two-ion rocking mode frequency}
The out-of-phase normal modes of a two-ion crystal are obtained by Taylor-expanding the potential of eq. (\ref{eq:Hr}) around the equilibrium positions of the ions.

In case of $\omega_y>\omega_x$, the ion crystal is oriented along the $x$-direction with $r_0=|\bm{r}|/2$ the equilibrium distance from the trap center. In the absence of micromotion at the ions' position ($q=0$), the Hamiltonian of the relative motion then becomes
\[
H_r=-\frac{\hbar^2}{2\mu}\nabla_r^2+\frac{1}{2}\mu(\omega_{s}^2u_x^2+\omega_{r}^2u_y^2) \]
where $u_x$, $u_y$ are the radial coordinates of the relative motion, $\omega_{s}=\sqrt{3}\omega_x$ is the stretch mode frequency and  $\omega_{r} = \sqrt{\omega_y^2-\omega_x^2}$ is the rocking mode frequency. If micromotion modifies the Coulomb potential ($q\neq 0$), the stretch mode frequency does not change but the rocking mode frequency is modified to
\begin{equation}
\omega_{r}^\prime=\sqrt{\omega_y^2-\omega_x^2(1+\frac{3}{2}q^2)} + {\cal{O}}(q^4). \label{eq:frockmod}
\end{equation}
The rocking mode potential becomes quartic when $\omega_{r}^\prime=0$ which happens at a critical normal mode splitting $\Delta\omega=\omega_y-\omega_x$ with
\begin{equation}
\Delta\omega=\frac{3}{4}q^2\omega_\perp \label{eq:frockcrit}
\end{equation}
where $\omega_\perp=\sqrt{(\omega_x^2+\omega_y^2)/2}$.
For $q=0.2$ and $\omega_\perp=(2\pi)\,\SI{1}{\mega\hertz}$, this would correspond to $\Delta\omega=(2\pi)\,\SI{30}{\kilo\hertz}$. In the absence of the quartic term, this would correspond to a critical rocking mode frequency of $\omega_{r}^{(c)}=\sqrt{\omega_y^2-\omega_x^2}=(2\pi)\,\SI{240}{\kilo\hertz}$. As the curvature of the time-averaged Coulomb potential (\ref{eq:AveragedCoulombPot}) in the radial direction at $\theta=\pm\pi/4$ equals the one at $\theta=0$ except for the opposite sign factor, the argument shows that, for degenerate frequencies $\omega_x=\omega_y$, the ion crystal would be trapped with its axis oriented under $\theta=\pm\pi/4$ and have a rocking mode frequency equal to $\SI{240}{\kilo\hertz}$.

\subsection{One-dimensional description of the exchange process}
As the confinement in the radial direction is much stiffer than along the direction normal to the crystal, we will neglect insignificant changes of the ion distance and small dispersive frequency shifts caused by cross mode coupling and focus only on the relative ion motion normal to the crystal axis. This leads to a one-dimensional description of the exchange process in terms of the crystal orientation. A good starting point for this is to rewrite the Hamiltonian that describes the relative motion in a harmonic trap in polar coordinates. If we assume that radial and angular dynamics completely decouple, then the angular part of the Hamiltonian is given by
\begin{equation}
H_\theta=-\frac{\hbar^2}{4mr_0^2}\frac{\partial^2}{\partial\theta^2} + mr_0^2(A\sin^2\theta+B\cos^2(2\theta))\label{eq:H_onedim}
\end{equation}
where $A=\omega_y^2-\omega_x^2$ and $B=\frac{3}{8}q^2\omega_\perp^2$. The part proportional to $B$ describes the micromotion-induced modification of the Coulomb interaction. Note that the magnitude of $A$ is equal to the square of the rocking mode frequency $\omega_r$ that one would calculate when using the secular approximation. The relative strength $A/B$ can be controlled by keeping the rf-voltage fixed and time-varying the dc-voltage that lifts the radial mode frequency degeneracy. For small dc-voltages, $A$ changes linearly with the voltage whereas $B$ remains constant.  
\begin{figure}[t]
\includegraphics[width=.5\textwidth]{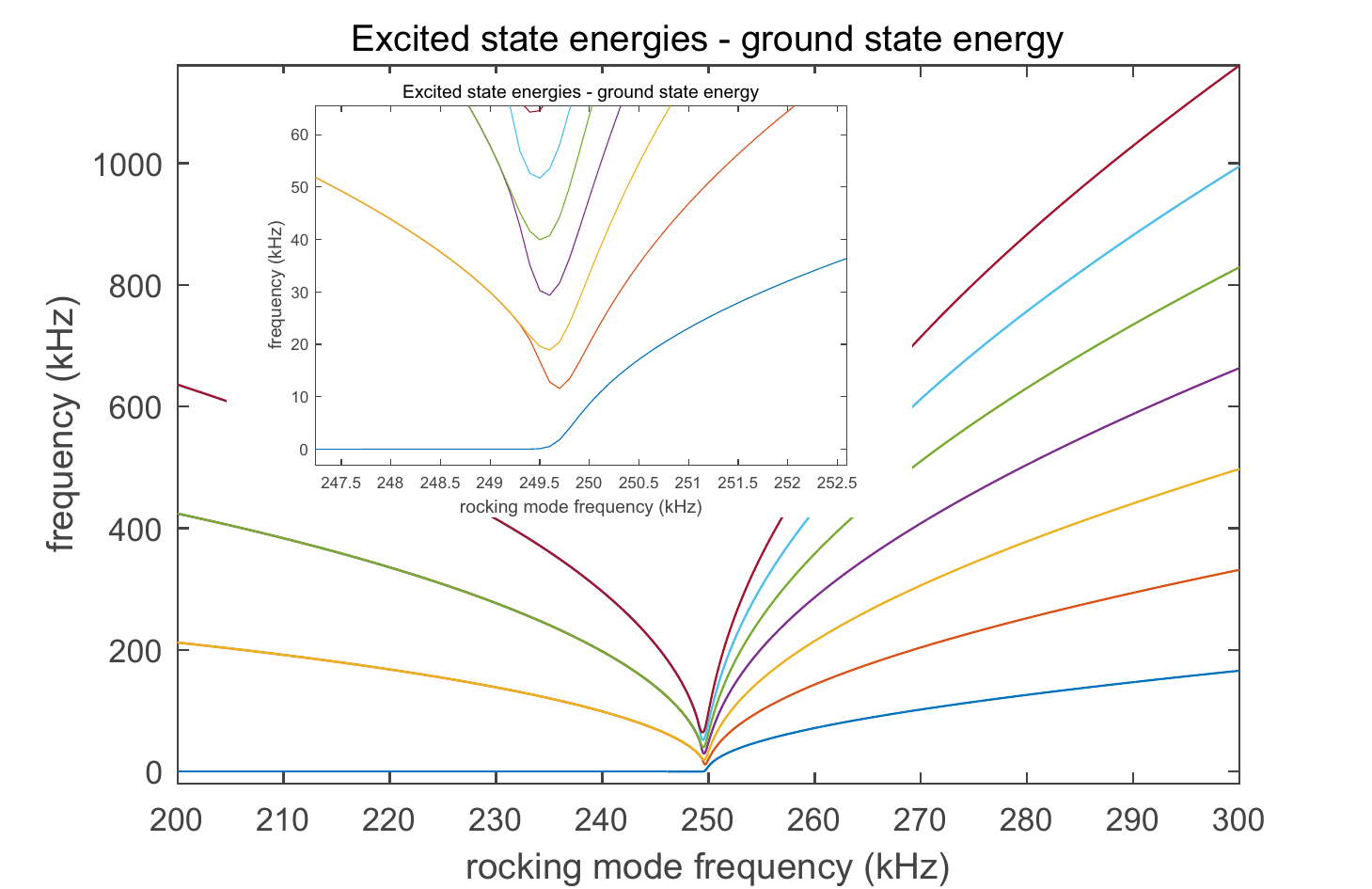}
\caption{\label{fig:EnergyLevels2} Energies of the lowest excited states during the voltage ramp that converts a harmonic potential (right-hand side) into a double well potential (left-hand side) obtained by diagonalization of the Hamiltonian of eq.~(\ref{eq:H_onedim}) for $q=0.2$ and $\omega_\perp=(2\pi)\,\SI{1}{\mega\hertz}$.} 
\end{figure}

Figure~\ref{fig:EnergyLevels2} shows the energy of excited states above the ground state energy as a function of $\omega_r/(2\pi)$ for $q=0.2$ and $\omega_\perp=(2\pi)\,\SI{1}{\mega\hertz}$. It can be seen that for $\omega_r/(2\pi)>\SI{250}{\kilo\hertz}$, the energy levels are approximately equidistant as expected for a harmonic oscillator. At about $\omega_r/(2\pi)=\SI{250}{\kilo\hertz}$, the potential becomes quartic before splitting into a double-well for $\omega_r/(2\pi)<\SI{250}{\kilo\hertz}$. In this regime, the energy levels become doubly degenerate corresponding to states where the ion is an an even or odd superposition of being in one or the other well. The gap separating the ground state of the lowest excited state of the same symmetry is always bigger than \SI{10}{\kilo\hertz}.
\begin{figure}[t]
\includegraphics[width=.5\textwidth]{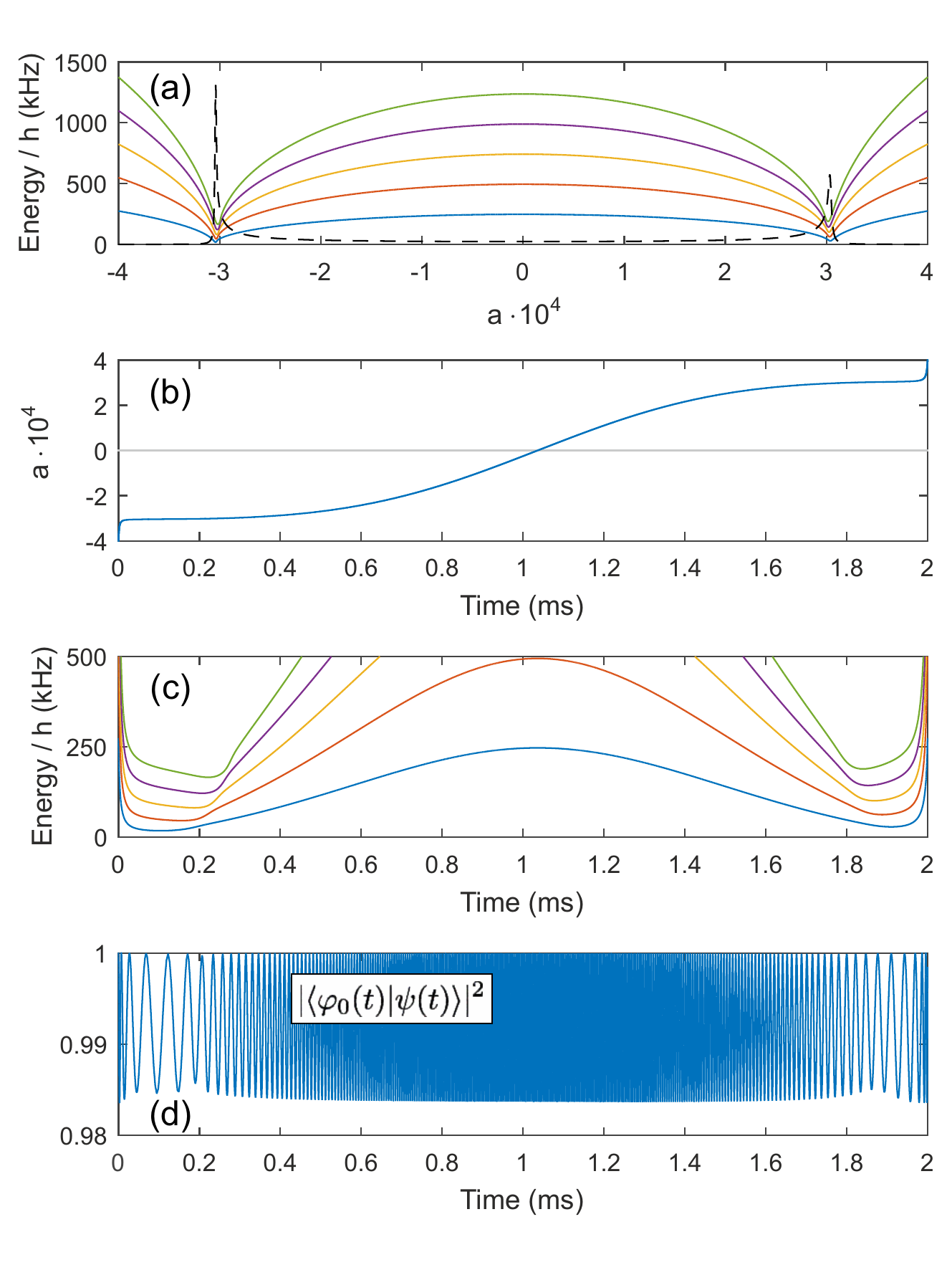}
\caption{\label{fig:splittingdynamics} Time dynamics of the single-to-double well splitting and subsequent recombination with an optimized frequency ramp of 2~ms duration. (a) Energies of excited states vs.\ $a$--parameter. The dashed line indicates the adiabaticity parameter of eq.~(\ref{eq:adiabaticity}) in arbitrary units. (b) $a$--parameter vs.\ time. (c) Energies of excited states vs.\ time. (d). Overlap squared of the time-evolved state with the ground state of the potential. For further details, see main text.}
\end{figure}

\subsection{Numerical simulation of the swapping process} 
The experiment could be carried out with a fermionic ion species like $^{40}$Ca$^+$ for which ground state cooling, and Bell state generation by entangling interactions are routine operations in experiments processing quantum information \cite{Schindler:2013}. 
We have carried out a simulation of the swapping process by numerically solving eq.~(\ref{eq:H_onedim}) for realistic parameters. We assume that the rf-trapping is achieved in a linear trap with a drive frequency of $\Omega_\text{rf}=(2\pi)\,\SI{20}{\mega\hertz}$ and $q=0.2$, and that the ions are axially confined by a static potential resulting in an oscillation frequency of $\omega_z=(2\pi)\,\SI{1.4}{\mega\hertz}$. In the absence of a dc-voltage applied to the rf-ground electrodes, the transverse oscillation frequency is then approximately given by $\omega_\perp=\frac{\Omega_\text{rf}}{2}\sqrt{q^2/2-a_z/2}$, where $a_z=(2\omega_z/\Omega_\text{rf})^2$ accounts for the deconfining effect of the axial quadrupole potential, so that $\omega_\perp\approx(2\pi)\,\SI{1}{\mega\hertz}$. A two-ion crystal will align in the radial plane with an ion distance of $2r_0\approx 5.6\mu$m. Ramping the dc-voltage from negative to positive values is equivalent to sweeping the trap's $a$-parameter which modifies the radial trapping frequencies to 
\begin{eqnarray}
\omega_x &=& \frac{\Omega_\text{rf}}{2}\sqrt{q^2/2-a_z/2+a}\\
\omega_y &=& \frac{\Omega_\text{rf}}{2}\sqrt{q^2/2-a_z/2-a}.
\end{eqnarray}  

For the simulation of the swapping process of a fermionic pair of particles, we could represent the Hamiltonian of eq.~(\ref{eq:H_onedim}) in a truncated basis of angular momentum eigenstates $\sim e^{in\theta}$, where $n = 2m-1$ with $m = -N, \ldots -1,0,1\ldots N$. However, if we start the simulation with $\omega_y>\omega_x$ and the rotor in the ground state of the potential, we can further restrict the state space by expanding the wave function and the Hamiltonian in $\psi_n\sim\cos(n\theta)$ with $n=1,3,\ldots 2N-1$.
In a first step, we calculate the energies $E_n(a)$ and eigenstates $\phi_{n}(\theta,a)$ by diagonalizing the Hamiltonian as a function of $a$ [see Fig.~\figref{fig:splittingdynamics}{a}]. We use this information to calculate an adiabaticity parameter for the ground state
\begin{equation}
\gamma(a) = \sum_{n>0}\frac{|\langle d \phi_n(\theta,a)/da|\phi_0(\theta,a)\rangle|}{E_n(a)-E_0(a)}.
\label{eq:adiabaticity}
\end{equation}
We use this parameter, shown as a dashed line in Fig.~\figref{fig:splittingdynamics}{a}, to convert a linear ramp of $a$ with time $t$ into the ramp $a(t)$ shown in Fig.~\figref{fig:splittingdynamics}{b} by scaling the time axis according to $dt(a)\rightarrow \gamma(a)dt(a)$. This procedure slows down the change in $a$ for the critical values where the single well changes into a double well potential. 
Then we adjust the time such that the $a$-parameter is ramped from $-4\cdot 10^{-4}$ to $+4\cdot 10^{-4}$ within 2 ms. Figure~\figref{fig:splittingdynamics}{c} shows the resulting energies of the eigenstates as a function of time. Finally, Fig.~\figref{fig:splittingdynamics}{d} demonstrates that the overlap of the time-evolved state $\psi(t)$ always has an overlap with the ground state bigger than 98\%. This demonstrates that the exchange process can be accomplished nearly adiabatically on realistic time scales for which decoherence effects are expected to be low.

\subsection{Effect of additional static fields: geometric and dynamical phases}
While the protocol is meant to detect the exchange phase $\varphi_\text{ex}$, the actually observed phase will always be the sum of $\varphi_\text{ex}$ and other dynamical or geometric phases caused by further interactions that were not included in the simple model. In order to anambiguously measure $\varphi_\text{ex}$, these additional phases need to be either very small, or they will have to be independently determined in additional control experiments. In the following, we will discuss a number of different interactions:

A non-zero magnetic field will give rise to an Aharonov-Bohm phase $\phi_{AB}=qB\pi r_0^2/\hbar$ where $r_0$ is the radius of the circle on which the ions move and $B$ is the strength of the magntic field component normal to the circle. For a field of $B=4$ Gauss normal to a circle with $r_0=2.5\mu$m, $\phi_{AB}\approx (2\pi)\,1.9$, i.e., the detected signal would be shifted by about two interference fringes. This effect could  however be easily calibrated. Moreover, it could be strongly reduced by suitably orientating the magnetic field direction.

The protocol has a high spatial and temporal symmetry that in principle prevents dynamical phases entering the measured signal. This symmetry might however by broken by spatially inhomogeneous fields. While phases caused by dc- or ac-magnetic field gradients are negligibly small, residual static electric stray fields, resulting in quadrupole potentials that are not aligned with the direction of the trap's normal modes of motion, are expected to have a significant impact on the measurement outcome. In the worst case, this effect might prevent the tranformation of the harmonic potential into a double-well potential. However, ion traps can be constructed that allow for full control of the static quadrupole potential such that residual stray field gradients could be compensated. In case of imperfect compensation, the stray fields will still add an additional phase factor to the measured signal. 

In order to analyze the impact of stray field gradients, we assume an quadrupolar stray electric potential with major axes that are rotated by $\pi/4$ with respect to dc quadrupole potential of the trap. In this potential, one part of the superposition state would be transported over a potential whereas the one part would pass by a potential valley; for this reason, a dynamical phase $\varphi_s$ would result. Such a potential would add an additional term $V_s(\theta)=mr_0^2A^\prime\sin^2(\theta+\pi/4)$ to the Hamiltonian of eq.~(\ref{eq:H_onedim}). Under the assumption that $V_s$ is tiny compared to the trapping potential, the resulting dynamical phase is approximately given by $\varphi_s=\int dt (V_s(\theta_{min}(t)-V_s(-\theta_{\text{min}}(t))/\hbar$. Here, we assumed a classical adiabatic transport starting at $\theta=0$ with $\theta_{\text{min}}(t)$ being the potential minimum of the trapping potential for $\theta\in[0,\pi/2]$. For the parameters of the transport process depicted in Fig.~ \ref{fig:splittingdynamics} and assuming that $A^\prime=8\cdot 10^8\,$s$^{-2}$, a very large dynamical phase $\varphi_s\approx 130\pi$ would result. We expect that this phase can be significantly reduced by optimizing transport process. A quick modification of the transport of Fig.~\ref{fig:splittingdynamics} led already to a reduction $\varphi_s\approx 10\pi$ at the expense of only 1\% loss of population to higher excited states. For a calibration of $\varphi_s$, we note that running the transport protocol forward and backwards followed by a measurement of the motional state would detect $2\varphi_s$ modulo $2\pi$. A complete characterization of $\varphi_s$ should be achievable by similar measurements after partial forward and backward transport.

The value of $A^\prime$ is more than two orders of magnitude smaller than the one observed in a recent experiment at IQOQI Innsbruck in a trap with an ion-to-electrode distance of about 500~$\mu$m. However, as stray charges on the electrodes or contact potentials give rise to potential curvatures that scale with the inverse square of the ion-to-electrode distance, the curvature could be reduced by an order of magnitude using a bigger trap and by preventing the trap electrodes from the being covered by the neutral atom beam used for loading the trap. Moreover, the stray potential curvatures are observed to be very stable over the course of many months. This opens up the possibility of detecting and compensating them with a set of dedicated compensation electrodes, thus reducing $A^\prime$ by a large amount.

\subsection{Sources of decoherence}

Decoherence of the electronic part of the wave function is expected to be completely negligible for the quantum states considered in this paper. The motional state might however decohere due to fluctuating electric stray fields. Fortunately, motional decoherence of the center-of-mass motion of the ion crystal would not affect the performance of the protocol. Motional decoherence of the relative motion is normally much weaker as compared to the one affecting the center-of-mass motion due to the small size of the ion crystal as compared to the distance to the nearest trap electrodes. 
Therefore, we don't expect the motional state of the quantum rotor, i.e. the angular motion and the stretch motion in the direction of the orientation of the rotor to be changeed by motiona heating within the time it takes to complete the protocol. Also, we are not aware of any mechanism that could give rise to fluctuating geometric phases that could dephase the state of a rotor in a superposition of two orientations on the experimentally relevant time scale.

While the quantum rotor mode is spectrally well separated from the other normal modes of motion, dispersive cross-Kerr-like couplings to other modes of relative motion can shift its frequency if the ions are confined in a trapping potential with near-degenerate frequencies \cite{Roos:2008a}. This effect can, however, be overcome by ground-state cooling the other two modes of relative motion. 

Temporal fluctuations of experimental control parameters could also give rise to effects similar to decoherence. In this respect, the most critical parameter is the control over the dc-voltage used to split the single-well potential into a double well which would require a relative stability of about $10^{-3}$ which can be realized with programmable precision voltage sources and a stabilization of the rf-power fed into the ion trap.

\subsection{Isotopic effects}
Trapped ion experiments could be carried out with ion species  (e.g., Ca$^+$ or Yb$^+$)  for which coherent manipulation of both bosonic and fermionic isotopes has been demonstrated in experiments. Because of the different isotope masses, however, it will not be possible to tune control parameters such that both trapping frequencies and ion distance stay the same. Therefore, dynamical phases arising from the ion transport will likely be different and require independent calibration.

\bibliographystyle{apsrev4-1}
\bibliography{references}

\begin{thebibliography}{67}
\bibitem [{Par()}]{ParityOperator}%
  \BibitemOpen
  \bibinfo {note} {The spin parity operator yields $1$ if the two atoms have
  same spin state, $-1$ if they have opposite spin state.}\BibitemShut {Stop}%
\end{thebibliography}

\end{document}